\documentclass[12pt]{article}
\usepackage[english]{babel}
\usepackage[latin1]{inputenc}
\usepackage{color}

\usepackage[intlimits]{amsmath}
\usepackage{amssymb}

\usepackage{url}
\usepackage{graphicx}

\numberwithin{equation}{section}

\providecommand{\unit}[1]{\ensuremath{\mathrm{#1}}}
\providecommand{\usk}{\ensuremath{\,}}

\begin{document}

\date{\mbox{ }}

\title{
{\normalsize
DESY 09-164\hfill\mbox{}\\
TUM-HEP 743/09\hfill\mbox{}\\
December 2009\hfill\mbox{}\\}
\vspace{1.5cm} 
\bf Neutrino Signals from Dark Matter Decay\\[8mm]}

\author{Laura Covi$^a$, Michael Grefe$^a$, Alejandro Ibarra$^b$, David Tran$^b$\\[2mm]
{\normalsize\it a Deutsches Elektronen-Synchrotron DESY, Hamburg}\\[-0.05cm]
{\it\normalsize Notkestra\ss{}e 85, 22603 Hamburg, Germany}\\[2mm]
{\normalsize\it b Physik-Department T30d, Technische Universit\"at M\"unchen,}\\[-0.05cm]
{\it\normalsize James-Franck-Stra\ss{}e, 85748 Garching, Germany}
}
\maketitle

\thispagestyle{empty}

\begin{abstract}
\noindent
We investigate different neutrino signals from the decay of dark matter particles to determine the prospects 
for their detection, and more specifically if any spectral signature can be disentangled from the background 
in present and future neutrino observatories. If detected, such a signal could bring an independent 
confirmation of the dark matter interpretation of the dramatic rise in the positron fraction above 10\usk GeV 
recently observed by the PAMELA satellite experiment and offer the possibility of distinguishing between 
astrophysical sources and dark matter decay or annihilation. In combination with other signals, it may also 
be possible to distinguish among different dark matter decay channels. 
\end{abstract}

\newpage

\section{Introduction}

While the existence of dark matter is now firmly established~\cite{DM}, little is known about the 
properties of the particles that make up the dark matter, including their lifetime. The dark matter 
particles are often assumed to be perfectly stable as the result of a symmetry, \textit{e.g.} $R$-parity 
in supersymmetric models. However, from the gravitational evidence for the existence of dark matter we can 
only infer directly that the dark matter has to be stable on timescales comparable to the age of the 
Universe. Measurements of cosmic-ray antimatter, gamma rays and neutrinos, on the other hand, typically 
impose much more stringent constraints on the dark matter lifetime in the respective decay 
channels~\cite{lifetime}. 

As with baryonic matter itself, there are good reasons to consider the case of unstable dark matter. From 
the theoretical point of view, we expect at least gravity to violate any global symmetry (and in some cases 
the symmetry breaking takes place at a lower scale, as \textit{e.g.} in models of Grand 
Unification~\cite{GUT-decay}). Therefore, we can expect the presence of at least non-renormalisable 
operators in the theory allowing for dark matter decay. In other cases, the dark matter parity symmetry 
may be only approximate from the start or spontaneously broken, as it occurs in some models of $R$-parity 
breaking~\cite{R-parity}. Alternatively, the coupling involved in the decay may be very strongly suppressed, 
as in the case of a tiny kinetic mixing between visible sector and hidden sector~\cite{kin-mix}. In those 
cases it is natural to expect a very long lifetime for the dark matter particle, which may exceed the age of 
the Universe by many orders of magnitude. Nevertheless, even for such extremely long lifetimes the decay 
signals may be in the observable 
range~\cite{GUT-decay,R-parity,kin-mix,DM-decay,Ibarra:2009dr,Covi:2008jy,neutrinos,leptophilic,PAMELA-decay}. 

Another important clue in relation to decaying dark matter is the observation of several excesses in the 
fluxes of cosmic rays in the energy region above a few GeV. Namely, the PAMELA experiment observed a steep 
rise in the positron fraction extending up to at least 100\usk GeV~\cite{Adriani:2008zr}. Furthermore, the 
experiments Fermi LAT, H.E.S.S. and ATIC measured the total $e^+ e^-$ flux, finding that this flux is harder 
than expected and falls off steeply above 1 TeV~\cite{FermiHessAtic}. Such signals can be well explained by 
the decay of dark matter particles in the Galactic halo if the decays are sufficiently ``leptophilic,'' such 
that one may avoid the overproduction of antiprotons. In fact, the observed antiproton flux is in complete 
agreement with the flux expected from production by spallation of cosmic rays~\cite{Adriani:2008zq}. 
Regarding the mass of the decaying particle, the PAMELA data require it to be larger than 200\usk GeV since the 
rise in the positron fraction extends up to at least 100\usk GeV, while the Fermi and H.E.S.S. measurements of 
the total $e^+ e^-$ flux indicate a mass of a few TeV, depending on the background~\cite{Hooper:2009cs} and 
the particular decay modes. The electron flux from dark matter decay required to fit the data points towards 
a lifetime of around $10^{26}$\usk s in all cases~\cite{Ibarra:2009dr}.

Similar signatures can also be generated by annihilating dark matter, but in this case some tension with 
the constraints coming from the measurement of radio emissions from the centre of the Galaxy~\cite{Radio} 
and from inverse Compton scattering with starlight and the cosmic microwave background~\cite{IC} is present. 
Note that astrophysical explanations for the positron excess have also been put forward~\cite{astro-positron}, 
the most popular being one or several nearby pulsars as additional sources of electrons and positrons. 

The decay of dark matter particles is an interesting and viable explanation of the observed electron 
anomalies, and it is worthwhile trying to confirm or exclude this possibility by a complementary examination 
of other indirect detection channels, in particular neutrinos~\cite{Covi:2008jy,neutrinos}. Neutrinos, from 
this perspective, have two clear advantages. Firstly, they are unaffected by magnetic fields and thus, like 
photons, allow to reconstruct the direction of their origin; therefore they would offer a clear way to 
distinguish between the cases of annihilating and decaying dark matter, as well as pulsar interpretations 
of the signal (even assuming that the pulsars produce also neutrinos in the energy range considered). 
Secondly, they are typically produced along with or from the decay of the charged leptons in many 
``leptophilic'' decaying dark matter models~\cite{leptophilic}. In such cases, therefore, the flux of neutrinos 
is correlated with the other cosmic-ray signals, and their spectrum may give direct information on the dark 
matter decay channel. In particular, choosing the mass and lifetime of the dark matter particle such as to
yield a good agreement with the PAMELA positron excess, one can directly predict the rates for the 
corresponding neutrino signal and look for it in present and future experiments. These are the two advantages 
we will try to exploit in this paper. 

On the other hand, neutrinos also suffer from some clear disadvantages with respect to other indirect 
detection channels: The large atmospheric neutrino background makes it difficult to disentangle any signal 
up to TeV energies for the lifetimes indicated by the cosmic-ray anomalies mentioned above. A further 
disadvantage is the necessity of very large detectors to measure the comparably small neutrino fluxes 
expected from dark matter decay. Fortunately, new large neutrino detectors, namely IceCube and possibly 
KM3NeT, will become fully operational in the near future and may allow to detect even the small signals we 
discuss here.

This paper is organised as follows: In section~\ref{Fluxes} we will discuss the neutrino flux expected from 
decaying dark matter in our Galaxy and compare it with the one from dark matter annihilation in order to 
discuss the best strategy for the detection of the signal in these two cases. In section~\ref{Spectra}, we 
will present the spectral signatures for a number of different dark matter decay modes. In section~\ref{Bounds} 
we will give the present bounds from neutrino experiments and the expected rates for present and future 
neutrino detectors. We will also discuss the prospects for distinguishing between different neutrino spectra 
in case a signal is detected. We will finally present our conclusions in section~\ref{Conclusions}.

\section{Neutrino Fluxes}
\label{Fluxes}

We concentrate in this section on the neutrino flux expected from the dark matter in the Milky Way halo, 
since on one hand it is the dominant source, and on the other hand it has a nontrivial directionality that 
may be exploited, as in the case of gamma rays~\cite{gamma-direction}, to disentangle the different 
hypotheses of dark matter decay versus annihilation. In addition, an isotropic extragalactic component is 
expected from unresolved cosmological sources, which in the case of decaying dark matter is of the same order 
of magnitude as the halo contribution, so that it may increase the overall signal by a factor of two or so. 
This extragalactic component is expected to be negligible in the case of dark matter annihilation.

\paragraph{Decaying Dark Matter}

For the case of decaying dark matter particles in the halo, the differential flux of neutrinos is given by 
the following integral along the line of sight:
\begin{equation}
 \frac{dJ_{\text{halo}}}{dE}=\frac{1}{4\pi\,\tau_{\text{DM}}\,m_{\text{DM}}}\,\frac{dN_{\nu}}{dE}\int_{\text{l.o.s.}}\rho_{\text{halo}}(\vec{l})\,d\vec{l}\;,
\end{equation}
where $\tau_{\text{DM}}$ and $m_{\text{DM}}$ are the lifetime and the mass of the decaying particle, 
$\frac{dN_{\nu}}{dE}$ is the neutrino energy spectrum from the decay and $\rho_{\text{halo}}$ is the dark 
matter density in the halo. Adopting for $\rho_{\text{halo}}$ the NFW density profile, we obtain for the 
averaged full-sky flux 
\begin{align}
 \left\langle \frac{dJ_{\text{halo}}}{dE}\right\rangle  &=\frac{\rho_{\text{loc}}\, 
 (R_{\odot}+r_c) }{4\pi\,\tau_{\text{DM}}\,m_{\text{DM}}}\,\frac{dN_{\nu}}{dE}
 \left(  \operatorname{artanh}\left(\frac{1}{1+\frac{r_c}{R_{\odot}}}\right)
 - \frac{\ln(\frac{R_{\odot}}{r_c})}{\frac{r_c}{R_{\odot}}-1}
 - \frac{1}{2} \ln\left(2\,\frac{R_{\odot}}{r_c}+1\right) \right) \nonumber\\
 &=1.3\times 10^{-8}\usk(\unit{cm^2\usk s\usk sr})^{-1}\left( \frac{10^{26}\usk\unit{s}}{\tau_{\text{DM}}}\right) \left( \frac{1\usk\unit{TeV}}{m_{\text{DM}}}\right) \frac{dN_{\nu}}{dE}\,, 
\end{align}
with the local halo density $\rho_{\text{loc}}=0.3$\usk GeV\usk cm$^{-3}$, the solar distance from the 
Galactic centre $R_{\odot}=8.5$\usk kpc and $r_c=20$\usk kpc. The numerical result is only weakly dependent 
on the halo parameters and the profile.

The flux is inversely proportional to the product of the dark matter particle mass and lifetime. Thus, for a 
fixed lifetime the flux is inversely proportional to the dark matter mass $m_{\text{DM}}$ due to the lower 
number density of dark matter particles for higher masses.

\paragraph{Annihilating Dark Matter}

For an annihilating particle, the differential flux of neutrinos is instead given by the following integral 
along the line of sight: 
\begin{equation}
 \frac{dJ_{\text{halo}}}{dE}=\frac{\left\langle \sigma v\right\rangle _{\text{DM}}}{8\pi\, m_{\text{DM}}^2}\,\frac{dN_{\nu}}{dE}\int_{\text{l.o.s.}}\rho_{\text{halo}}^2(\vec{l})\,d\vec{l}\;,
\end{equation}
where $\left\langle \sigma v\right\rangle _{\text{DM}}$ and $m_{\text{DM}}$ are the dark matter annihilation 
cross-section and the dark matter mass, $\frac{dN_{\nu}}{dE}$ is the neutrino spectrum from annihilation 
instead of decay, and the line-of-sight integral, in this case, contains the square of the halo density. 
It is therefore clear that for the same halo profile the annihilating dark matter signal is strongly enhanced 
towards the centre of the Galaxy, especially for cuspy halo profiles.

\paragraph{Propagation}

After the neutrinos are produced in the decay or annihilation of dark matter particles, they travel in 
straight lines through the Galaxy, essentially without any interactions. The only modifications to the fluxes 
during this time are due to flavour oscillations~\cite{Strumia:2006db}. In fact, using the experimental 
best-fit values for the neutrino mixing angles, $\sin^2\theta_{12}=0.304$, $\sin^2\theta_{23}=0.5$ and 
$\sin^2\theta_{13}=0.01$~\cite{Schwetz:2008er}, and neglecting possible CP-violating effects, the neutrino 
oscillation probabilities in vacuum are given by 
\begin{equation}
 \begin{split}
 P(\nu_e\leftrightarrow\nu_e) &=0.56\,, \\
 P(\nu_e\leftrightarrow\nu_{\mu})=P(\nu_e\leftrightarrow\nu_{\tau}) &=0.22\,, \\
 P(\nu_{\mu}\leftrightarrow\nu_{\mu})=P(\nu_{\mu}\leftrightarrow\nu_{\tau})=P(\nu_{\tau}\leftrightarrow\nu_{\tau}) &=0.39\,.
 \end{split}
 \label{oscprob}
\end{equation}
Thus, a primary neutrino flux in a specific flavour is redistributed almost equally into all neutrino flavours 
during propagation and any flavour information is lost. On the other hand, this means that nearly the same 
signal is present in any flavour and may allow to choose the best channel for discovery according to the 
background and efficiency of the detector.

\subsection{Background Fluxes}

Let us now discuss the background for our neutrino signal. The main background for the observation of 
neutrinos in the GeV to TeV range are neutrinos produced in cosmic-ray interactions with the Earth's 
atmosphere. Here we use the atmospheric neutrino fluxes calculated by Honda \textit{et al.}~\cite{Honda:2006qj}. 
The theoretical uncertainty of these fluxes is estimated to be better than 25\usk\% in the GeV to TeV range, 
while the uncertainty in the ratio of the different flavours is significantly smaller. We extend the 
atmospheric neutrino fluxes to energies above 10\usk TeV using the slopes given by Volkova \textit{et 
al.}~\cite{Volkova:2001th}. 

Conventional electron and muon neutrinos are directly produced from pion and kaon decays. While electron 
neutrinos are practically unaffected by neutrino oscillations due to the large oscillation length, muon 
neutrinos, particularly at low energies, can be converted into tau neutrinos and provide the dominant tau 
neutrino background at energies below 1\usk TeV. The conversion probability of muon neutrinos into tau 
neutrinos is given by 
\begin{equation}
 P(\nu_{\mu}\rightarrow\nu_{\tau})\simeq\sin^2\left( 3.05\times 10^{-3}\,\frac{L\usk(\unit{km})}{E_\nu\usk(\unit{GeV})}\right).
\end{equation}
In this expression, $E_\nu$ is the neutrino energy and $L$ is their propagation length after being produced 
in the atmosphere, which is given by 
\begin{equation}
 L(\theta)=\sqrt{(R_{\oplus}\cos\theta)^2+2R_{\oplus}h+h^2}-R_{\oplus}\cos\theta\,,
\end{equation}
with $R_{\oplus}\simeq 6.4\times 10^3$ \usk km being the Earth's radius and $h\simeq 15$ \usk km the mean 
altitude at which atmospheric neutrinos are produced. 

In addition to the conventional atmospheric neutrino flux from pion and kaon decays there is a prompt neutrino 
flux from the decay of charmed particles that are also produced in cosmic-ray collisions with the atmosphere. 
The prompt neutrinos have a harder spectrum than the conventional ones and therefore dominate at higher 
energies (roughly above 10\usk TeV for electron neutrinos and above 100\usk TeV for muon neutrinos). Since 
these contributions are not well understood and in any case subdominant in the energy range that is of interest 
here, we neglect them in the present study. 
 
On the other hand, the prompt tau neutrinos start to dominate around 1\usk TeV (and at even smaller energies 
for downgoing neutrinos). Thus we include this contribution using the parametrisation~\cite{Pasquali:1998xf} 
\begin{equation}
 \log_{10}\left[ E^3\,\frac{dJ_{\nu_{\tau}}}{dE}\left/ \left( \frac{\unit{GeV}^2}{\unit{cm^2\usk s\usk sr}}\right) \right. \right] =-A+Bx-Cx^2-Dx^3,
\end{equation}
where $x=\log_{10}\left( E\usk(\unit{GeV})\right) $, $A=6.69$, $B=1.05$, $C=0.150$ and $D=-0.00820$. This 
parametrisation is valid in the energy range of 100\usk GeV up to 1\usk PeV. However, we point out that 
compared to the conventional atmospheric neutrino flux the prompt flux suffers from larger uncertainties. 

Other neutrino backgrounds in the considered energy range are neutrinos produced in cosmic-ray interactions 
with the solar corona~\cite{Ingelman:1996mj} and those produced in cosmic-ray interactions with the 
interstellar medium in the Milky Way~\cite{Athar:2004um}. While the former is subdominant in diffuse searches 
for all flavours~\cite{Covi:2008jy} and can be excluded from the analysis by excluding neutrinos from the 
direction of the Sun, the latter represents an irreducible, ill-understood neutrino background for searches in 
the Galactic disc direction. In fact, the flux of Galactic neutrinos is expected to become comparable to the 
atmospheric electron neutrino background for the Galactic disc direction and energies in the TeV range.

\subsection{General Detection Strategy and Use of Directionality}
\label{direction}

In view of the subdominant neutrino signals from dark matter decays it is important to devise strategies that 
reduce the background. In~\cite{Covi:2008jy} it was proposed to use directionality in order to reduce the 
background in the tau neutrino channel. This is possible since the tau neutrino background at low energies 
comes mainly from the muon neutrino oscillation and is therefore strongly suppressed in the zenith direction. 
Some of the authors of~\cite{neutrinos} propose instead to search for an enhanced muon neutrino signal only 
in the direction of the Galactic centre. However, taking into account the typically low neutrino event rates, 
it is not always the best strategy to optimise the signal-to-background ratio. Instead, the statistical 
significance $\sigma=S/\sqrt{B}$ (number of signal events divided by the square root of the number of 
background events) is a better measure for comparing different detection strategies. 

\begin{figure}
 \centering
 \includegraphics[scale=1.,bb=53 49 404 296,clip]{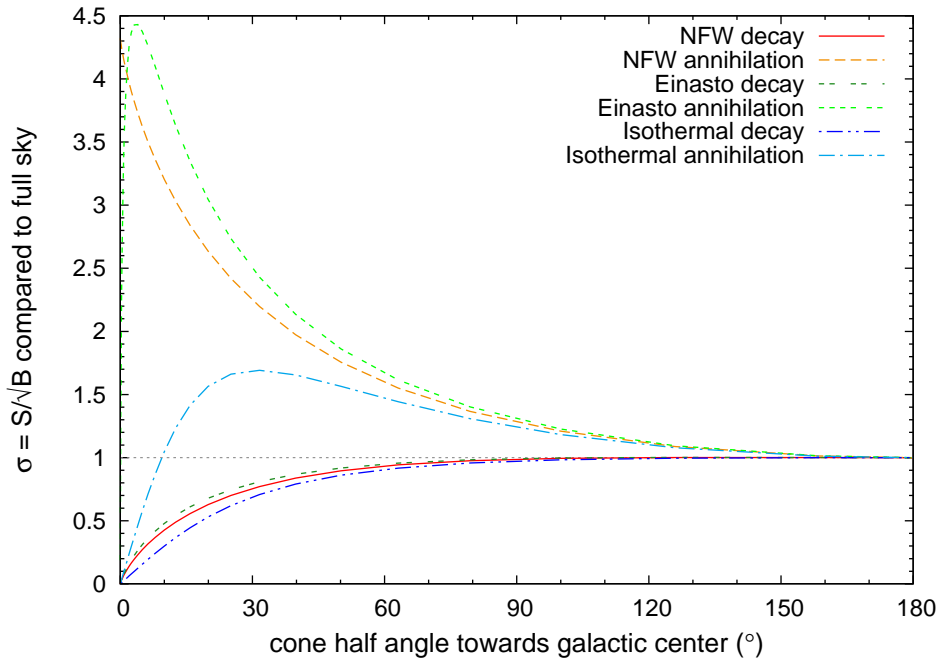}
 \caption{Significance of the signal as a function of the cone half angle towards the Galactic centre normalised 
 to the significance of the full-sky signal for annihilating/decaying dark matter depending on the different 
 density profiles.}
\label{GCdirection}
\end{figure}

In Figure~\ref{GCdirection} we show the significance of the signal as a function of the cone half angle around 
the Galactic centre normalised to the significance of the full-sky observation. Here we assume a background 
that (practically) does not depend on Galactic coordinates like atmospheric neutrinos and neglect the 
contribution of Galactic neutrinos, which is subdominant in the case of muon neutrinos. We see clearly that 
for annihilating dark matter the best way to detect the signal is indeed looking towards the Galactic centre: 
The cone half angle offering the best signal-to-square root of background ratio varies depending on the 
cuspiness of the profile, but it is always between $\sim0^\circ$ (NFW) and $30^\circ$ (isothermal). Note that 
in any case the gain of looking at the Galactic centre is not very large for a cored profile like the 
isothermal one. 

For the case of decaying dark matter on the other hand, the best strategy is to measure the full-sky signal 
and not concentrate on the region around the Galactic centre. In fact the gain coming from the enhanced dark 
matter density is counteracted by the smallness of the collecting area and so the significance of the signal 
goes quickly to zero as a function of the angle for any profile, even for cuspy profiles like the NFW profile. 
The observation of only a fraction of the sky around the Galactic centre direction leads to an increase in the 
signal-to-background ratio, but not of the significance. We therefore conclude that for decaying dark matter 
there is no advantage in looking only at the Galactic centre. The full-sky signal offers not only better 
statistics, but also a higher significance. 

Considering the directionality of the atmospheric background instead of the signal, another good strategy 
might be to exploit the fact that the flux from the zenith direction is (dependent on the energy) a few times 
smaller than from the horizontal direction. Assuming a signal that does not depend on the zenith angle, the 
observation of only a fraction of the sky around the zenith direction is again clearly leading to an increase 
in the signal-to-background ratio. Also in this case though, it turns out that the best value for the 
significance is achieved for a full-sky observation. 

We can therefore conclude that exploiting the directionality of the signal or background, apart from the case 
of specific flavours like the tau neutrino discussed in~\cite{Covi:2008jy}, is not promising for the first 
detection of decaying dark matter. The largest rate and significance is achieved for a full-sky search, and 
this is the option we will discuss in the following. On the other hand, directionality offers a clear way to 
disentangle decaying dark matter from either annihilating dark matter, where looking into the Galactic centre 
should give an increase in significance, or from point sources like dwarf galaxies, pulsars and supernova 
remnants.

\section{Neutrino and Muon Spectra}
\label{Spectra}

The neutrino spectra depend on the decay channel of the dark matter particle. The simplest possibility is a 
direct decay into two neutrinos for a scalar particle or into $Z^0\nu$ for a fermion. Then the resulting 
spectrum is just a monochromatic line for the Galactic signal and an integral of the redshifted line from the 
extragalactic signal (and a continuum contribution from the fragmentation of the $Z^0$ boson in the case of 
decay into $Z^0\nu$). So for this case we have the simple spectra 
\begin{equation}
 \begin{split}
  \frac{dN_{\nu}}{dE}\left(\text{DM} \rightarrow X \nu\right) \propto & \;\delta\left( E-\frac{m_\text{DM}}{2}\right) \\
  &+C_{\text{l.o.s.}}\left( 1 + \kappa \left( \frac{2\, E}{m_\text{DM}} \right)^3 \right)^{-1/2}
  \left( \frac{2\, E}{m_\text{DM}} \right)^{1/2} \Theta\left( E-\frac{m_\text{DM}}{2}\right) \\
  &\left( +\;\text{continuum}\right) ,
 \end{split}
\end{equation}
where $X = Z^0, \gamma, \nu$ and we have assumed here that the mass of $X$ is negligible. $C_{\text{l.o.s.}}$ is the ratio of the extragalactic and the Galactic signal, which is a number of 
order one given by
\begin{equation}
 C_{\text{l.o.s.}} = \frac{ \Omega_\text{DM}\, \rho_c}{H_0\, \Omega_M^{1/2} }\left(\; \int_{\text{l.o.s.}}\rho_{\text{halo}}(\vec{l})\,d\vec{l}\right) ^{-1} 
\end{equation}
and $\kappa=\Omega_\Lambda/\Omega_m\approx 3$ is the ratio of the dark energy density and the matter density 
in the Universe.

\begin{figure}
 \centering
 \includegraphics[scale=1.02,bb=50 70 396 293,clip]{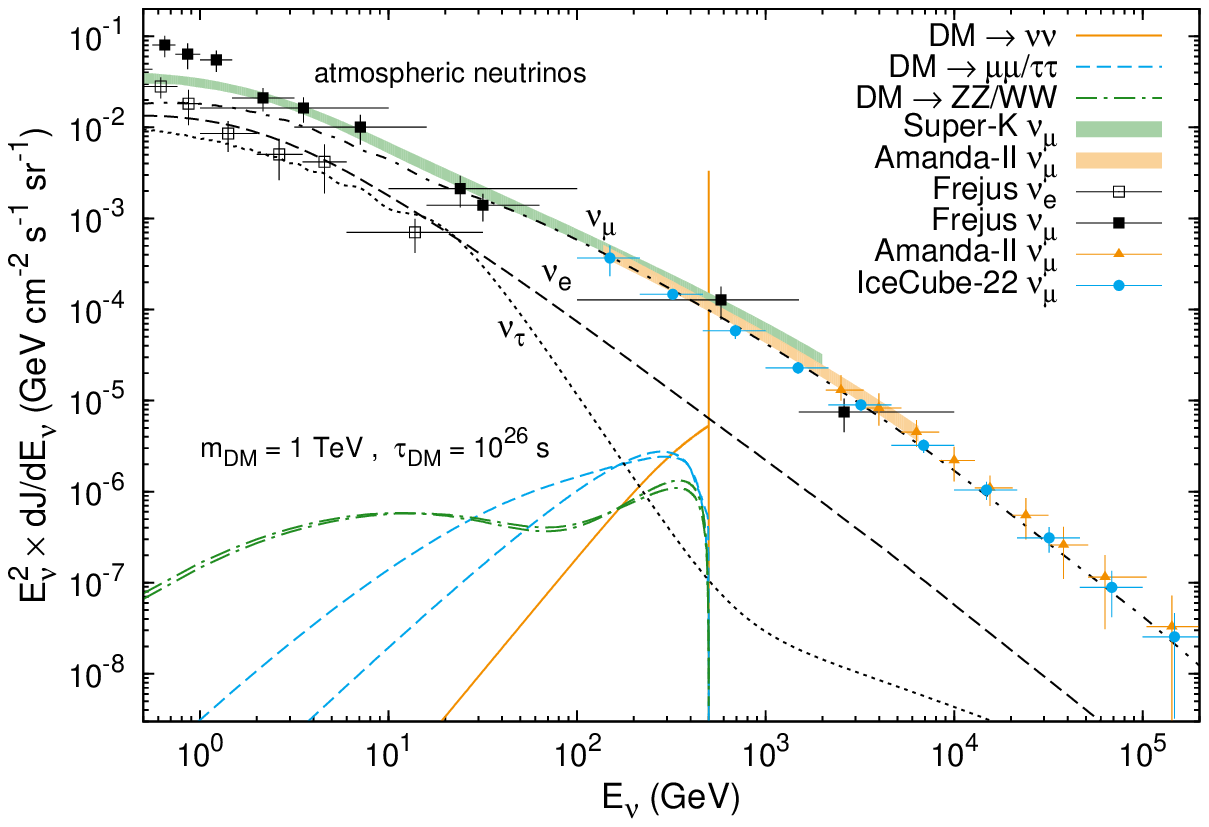}
 \includegraphics[scale=1.02,bb=50 58 396 293,clip]{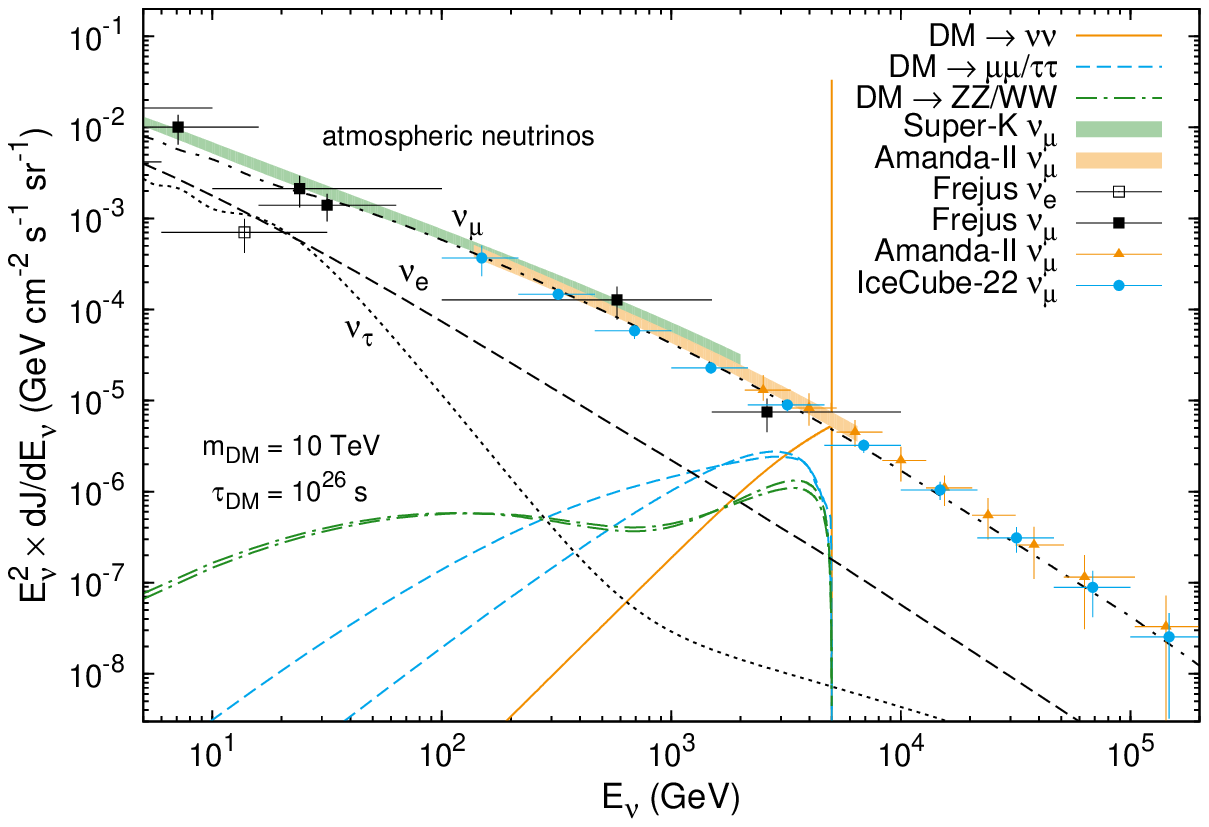}
 \caption{Neutrino spectra for different decay channels of a scalar dark matter candidate compared to the 
 expected background of atmospheric neutrinos from Honda \textit{et al.}~\cite{Honda:2006qj} and the data of 
 Fr\'ejus~\cite{Daum:1994bf}, Super-Kamiokande~\cite{GonzalezGarcia:2006ay}, 
 Amanda-II~\cite{Collaboration:2009nf} and IceCube~\cite{Chirkin:2009}. The flux is computed for a dark matter 
 mass of 1\usk TeV ({\it top}) or 10\usk TeV ({\it bottom}) and a lifetime of $10^{26}$\usk s. The line from 
 the two-body decay into $\nu\bar{\nu}$ and the extragalactic contribution to this decay spectrum is easy to 
 distinguish. The spectra from the decays of a dark matter candidate into $\mu^+\mu^-$, $\tau^+\tau^-$, 
 $Z^0Z^0$ or $W^\pm W^\mp$ are softer at the endpoint. The low-energy tail of these decay channels is due to 
 the muon/tau decay and $Z^0$/$W^\pm$ fragmentation. Due to the steeply falling atmospheric background the 
 signal-to-background ratio at the endpoint of the decay spectra increases significantly for larger dark 
 matter masses.}
\label{Spectra-plot}
\end{figure}

\begin{figure}
 \centering
 \includegraphics[scale=1.02,bb=50 70 396 293,clip]{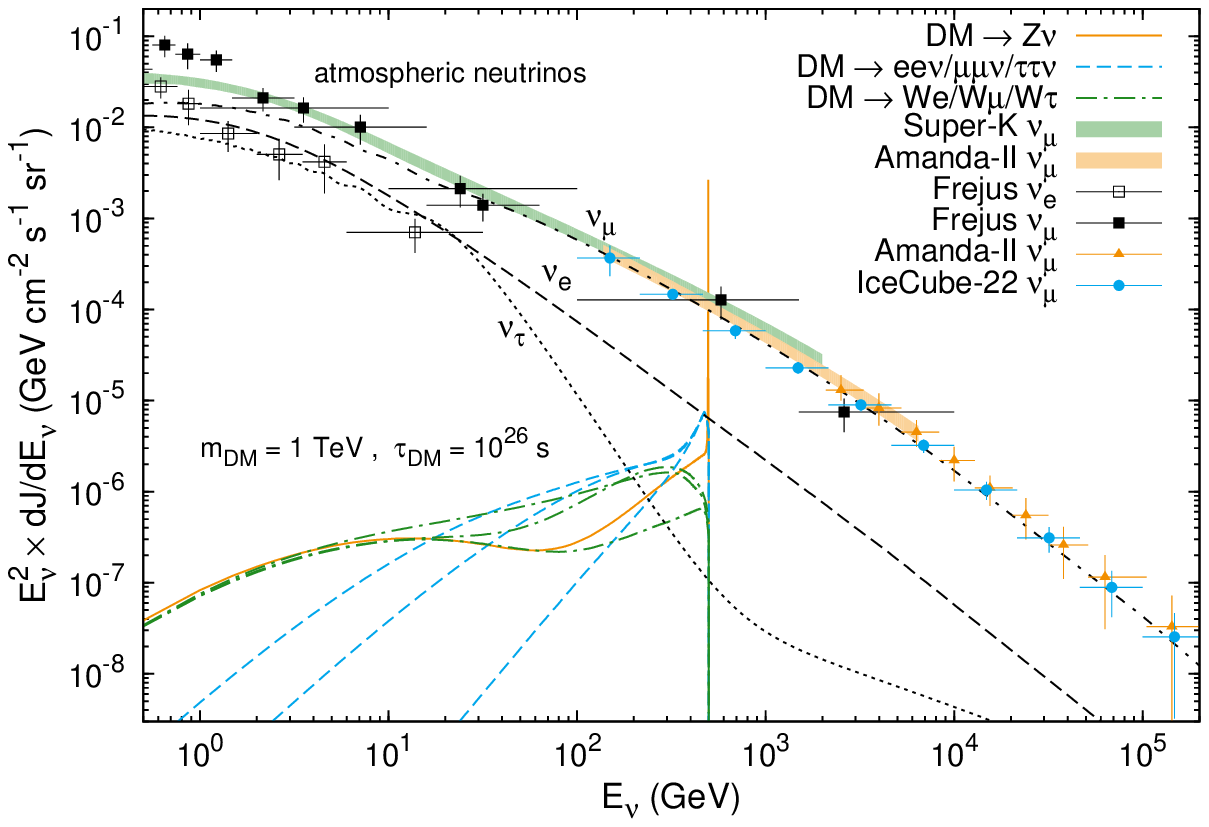}
 \includegraphics[scale=1.02,bb=50 58 396 293,clip]{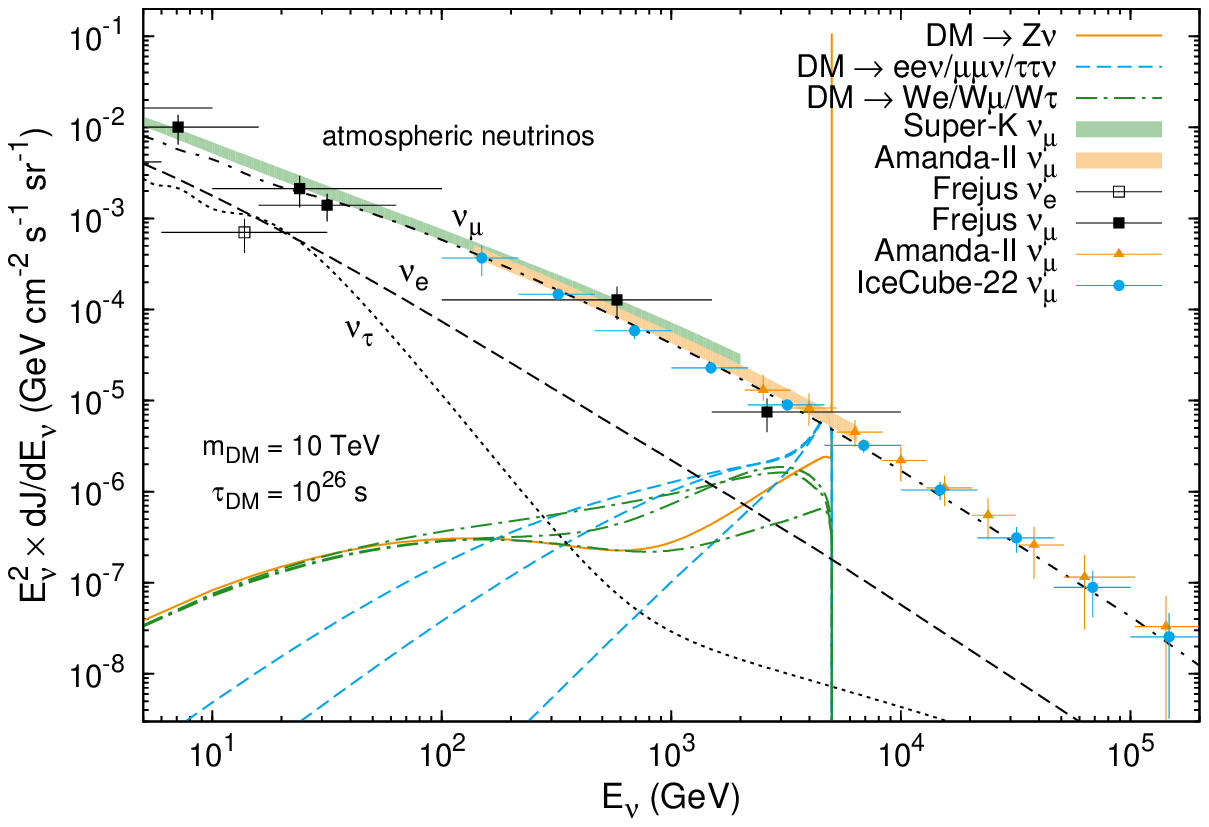}
 \caption{Neutrino spectra for different decay channels of a fermionic dark matter candidate compared to the 
 expected background of atmospheric neutrinos from Honda \textit{et al.}~\cite{Honda:2006qj} and the data of 
 Fr\'ejus~\cite{Daum:1994bf}, Super-Kamiokande~\cite{GonzalezGarcia:2006ay}, 
 Amanda-II~\cite{Collaboration:2009nf} and IceCube~\cite{Chirkin:2009}. The flux is computed for a dark 
 matter mass of 1\usk TeV ({\it top}) or 10\usk TeV ({\it bottom}) and a lifetime of $10^{26}$\usk s. The 
 pure line and three-body decays are easy to distinguish and correspond to the line contained in $Z^0\nu$ 
 and the decay into $e^+ e^-\nu$. Note that the low-energy tail of $Z^0\nu$ and of the other leptonic 
 three-body decays is due to the $Z^0$ fragmentation and muon/tau decay. Also shown are the cases of a pure 
 continuum spectrum coming from the decay into $W^\pm\ell^\mp$. Due to the steeply falling atmospheric 
 background the signal-to-background ratio at the endpoint of the decay spectra increases significantly for 
 larger dark matter masses.}
 \label{Spectra-plot2}
\end{figure}

Another characteristic spectrum is that obtained from a three-body decay into three leptons $\ell^+\ell^-\nu$, 
which has the familiar triangular shape when plotted on a logarithmic axis.\footnote{This holds for the most 
common scenarios, where the decay is mediated by a heavy scalar or a heavy vector boson. In both cases the 
Michel parameter $\rho$ is equal to 3/4, yielding the same neutrino energy spectrum.} In this case the 
expression for the extragalactic signal is more involved, but it still appears as a softer triangular shape 
which dominates at low energies, as can be seen from the change in slope for the three-body spectra in 
Figures~\ref{Spectra-plot} and~\ref{Spectra-plot2}. We show here as an example the spectrum of the decay of a 
fermionic dark matter particle mediated by a heavy scalar particle, corresponding to a scalar-type 4-fermion 
interaction. 

Finally, a continuous neutrino spectrum is generated by any heavy particle that decays into neutrinos, like 
the muon, or fragments into charged pions, like the electroweak gauge bosons or the tau lepton. We use here 
as examples of continuum neutrino spectra, the spectra arising from the decay of a scalar particle into 
longitudinal gauge bosons similar to the Higgs decay and from the decay of a fermionic chiral lepton into
$W^{\pm} \ell^{\mp}, Z^0 \nu$. We use PYTHIA 6.4~\cite{Sjostrand:2006za} to simulate the gauge boson 
fragmentation and the heavy leptonic decays.

The corresponding neutrino fluxes for these different types of spectra are shown in Figures~\ref{Spectra-plot} 
and~\ref{Spectra-plot2} for a scalar and a fermionic dark matter candidate, respectively, together with the 
expected atmospheric background and the data measured by the Fr\'ejus~\cite{Daum:1994bf}, 
Super-Kamiokande~\cite{GonzalezGarcia:2006ay}, AMANDA-II~\cite{Collaboration:2009nf} and 
IceCube~\cite{Chirkin:2009} experiments. We see that for a lifetime of the order of $10^{26}$\usk s, which is 
the order of magnitude suggested by the PAMELA excess~\cite{PAMELA-decay}, the signal always lies below the 
measured background of muon neutrinos. The best signal-to-background ratio is achieved for the high-energy 
end of the spectrum, which gives information about the mass scale of the decaying particle. 

The neutrino spectra shown in Figures~\ref{Spectra-plot} and~\ref{Spectra-plot2} look rather distinctive, and 
an interesting question is whether they can be disentangled in a neutrino detector. We have to consider that 
neutrino detectors do not really measure neutrinos directly, but the corresponding charged leptons or showers, 
produced in the interactions of neutrinos with the intervening matter.

\subsection{Neutrino Interactions}

Since we are interested in neutrino energies much larger than nucleon masses we only take into account deep 
inelastic neutrino--nucleon scattering. Neutrino--electron elastic scattering is subdominant in this energy 
range and will be neglected. 

The cross-sections for deep inelastic scattering of (anti)neutrinos off nucleons at rest are given by 
\begin{equation}
 \begin{split}
  \frac{d\sigma_{\text{CC/NC}}^{\nu\, p,n}(E_{\nu}, y)}{dy} &\simeq\frac{2\,m_{p,n}\,G_F^2}{\pi}\,E_{\nu}\left( a_{\text{CC/NC}}^{\nu\,p,n}+b_{\text{CC/NC}}^{\nu\,p,n}\,(1-y)^2\right) \\
  &\simeq 3.2\times 10^{-38}\usk\frac{\unit{cm^2}}{\unit{GeV}}\,E_{\nu}\left( a_{\text{CC/NC}}^{\nu\,p,n}+b_{\text{CC/NC}}^{\nu\,p,n}\,(1-y)^2\right) \\
 \end{split}
 \label{crosssection}
\end{equation}
with $a_{\text{CC}}^{\nu\,p,n}=0.15,\,0.25$, $b_{\text{CC}}^{\nu\,p,n}=0.04,\,0.06$ and 
$a_{\text{CC}}^{\bar{\nu}\,p,n}=b_{\text{CC}}^{\nu\,n,p}$, 
$b_{\text{CC}}^{\bar{\nu}\,p,n}=a_{\text{CC}}^{\nu\,n,p}$ for charged-current interactions, and 
$a_{\text{NC}}^{\nu\,p,n}=0.058,\,0.064$, $b_{\text{NC}}^{\nu\,p,n}=0.022,\,0.019$ and 
$a_{\text{NC}}^{\bar{\nu}\,p,n}=b_{\text{NC}}^{\nu\,p,n}$, 
$b_{\text{NC}}^{\bar{\nu}\,p,n}=a_{\text{NC}}^{\nu\,p,n}$ for neutral-current 
interactions~\cite{Strumia:2006db,Barger:2007xf}. The inelasticity $y$ is given by 
\begin{equation}
 y=1-\frac{E_\ell}{E_{\nu}}\qquad\text{or}\qquad y\simeq \frac{E_{\text{had}}}{E_{\nu}}\,,
\end{equation}
where $E_\ell$ is the energy of the generated lepton and $E_{\text{had}}$ is the energy of the generated 
hadronic shower. Equation~(\ref{crosssection}) holds only for neutrino energies up to the TeV region when 
the effect of the massive gauge boson propagators cannot be neglected anymore. For higher energies the 
cross-sections are overestimated. 

For the total neutrino--nucleon cross-sections one obtains 
\begin{equation}
 \begin{split}
  \sigma_{\text{CC/NC}}^{\nu\, p,n}(E_{\nu}) &\simeq\frac{2\,m_{p,n}G_F^2}{\pi}\,E_{\nu}\left( a_{\text{CC/NC}}^{\nu\,p,n}+\frac{1}{3}\,b_{\text{CC/NC}}^{\nu\,p,n}\right) .
 \end{split}
 \label{totalcross}
\end{equation}
As we can see, the total cross-section is proportional to the energy of the incoming neutrino in the 
considered energy range.

\subsection{Muon Neutrinos}

The charged-current deep inelastic scattering of a muon neutrino off a nucleus produces a hadronic shower and 
a muon. These track-like events can be clearly identified in Cherenkov detectors via the Cherenkov light cone 
of the relativistic muon.

\subsubsection{Through-going Muons}

Since muons are rather long-lived ($c\tau_\mu=658.650\usk$m), their range is only limited by energy loss 
during their passage through matter and not by their lifetime. Therefore Cherenkov detectors can also observe 
muons that are generated in the surrounding material of the detector. This effect enhances the effective 
detector area for high-energy muon neutrinos. 

The average rate of muon energy loss can be written as 
\begin{equation}
 -\frac{dE_{\mu}}{dx}=\alpha(E_{\mu})+\beta(E_{\mu})\,E_{\mu}\,,
 \label{energyloss}
\end{equation}
where $\alpha(E_{\mu})$ describes the ionisation energy loss and $\beta(E_{\mu})$ takes into account the 
energy loss due to radiative processes: $e^+e^-$ pair production, bremsstrahlung and photonuclear 
contributions. Both $\alpha(E_{\mu})$ and $\beta(E_{\mu})$ are slowly varying functions of the muon energy. 
As long as we can approximate $\alpha$ and $\beta$ as energy-independent, the average range after which the 
muon energy drops below a threshold energy $E_{\mu}^{th}$ is given by 
\begin{equation}
 R_{\mu}(E_{\mu},E_{\mu}^{\text{th}})=\frac{1}{\rho\,\beta}\,\ln\left[ \frac{\alpha+\beta E_{\mu}}{\alpha+\beta E_{\mu}^{\text{th}}}\right] , 
 \label{muonrange}
\end{equation}
where $\rho$ is the density of the medium. The relevant parameters for standard rock, water and ice are given 
in Table~\ref{materials}. The values of the density and the average proton-number-to-mass-number ratio are 
taken from~\cite{Lohmann:1985qg}. The muon energy loss parameters given in the table are best-fit values from 
the fit of equation~(\ref{muonrange}) to the tabulated data in~\cite{Lohmann:1985qg}. 
\begin{table}
 \centering
 \begin{tabular}{lcccc}
  \hline
  material & density (g/cm$^3$) & $\left\langle Z/A\right\rangle $ & $\alpha$ (GeV\usk cm$^2$/g) & $\beta$ (cm$^2$/g) \\
  \hline
  standard rock & 2.650 & 0.5 & $2.3\times 10^{-3}$ & $4.4\times 10^{-6}$ \\
  water & 1.000 & 0.55509 & $2.7\times 10^{-3}$ & $3.3\times 10^{-6}$ \\
  ice & 0.918 & 0.55509 & $2.7\times 10^{-3}$ & $3.3\times 10^{-6}$ \\
  \hline
 \end{tabular}
 \caption{Density, proton-number-to-mass-number ratio and approximate muon energy loss parameters for the 
 materials of interest in Cherenkov detectors.}
 \label{materials}
\end{table}
From equation~(\ref{muonrange}) we can determine the initial muon energy as a function of the final muon 
energy and the muon range: 
\begin{equation}
 E_{\mu}^0(E_{\mu})=E_{\mu}e^{\beta\rho r}+\frac{\alpha}{\beta}\left( e^{\beta\rho r}-1\right) .
 \label{muonenergy}
\end{equation}
In fact, equation~(\ref{energyloss}) does not account for the stochastic nature of radiative muon energy 
losses which start to dominate at TeV energies ($E>\alpha/\beta$), and therefore equation~(\ref{muonrange}) 
overestimates the muon range for large energies. 

The rate of muon neutrino induced through-going muon events is given by 
\begin{equation}
 \begin{split}
  \frac{dN}{dt}= &\int d\Omega\int_0^{\infty} dE_{\nu_{\mu}}\, \frac{dJ_{\nu_{\mu}}(E_{\nu_{\mu}},\theta,\phi)}{dE_{\nu_{\mu}}}\,A^{\text{eff}}_{\nu_{\mu}}(E_{\nu_{\mu}},\theta,\phi) \\
  = &\int d\Omega\int_{E_{\mu}^{\text{th}}}^{\infty} dE_{\nu_{\mu}}\int_{E_{\mu}^{\text{th}}}^{E_{\nu}}dE_{\mu}\, \frac{dJ_{\nu_{\mu}}(E_{\nu_{\mu}},\theta,\phi)}{dE_{\nu_{\mu}}}\left[ \frac{d\sigma_{\text{CC}}^{\nu p}(E_{\nu_{\mu}},E_{\mu})}{dE_{\mu}}\,n_p+(p\rightarrow n)\right] \\
  &\times R_{\mu}(E_{\mu},E_{\mu}^{\text{th}})\,A^{\text{eff}}_{\mu}(E_{\mu},\theta,\phi)\,e^{-\sigma^{\nu N}(E_{\nu_{\mu}})\,n_N\,L(\theta)}+(\nu\rightarrow\bar{\nu})\,,
 \end{split}
 \label{through-rate}
\end{equation}
where the number density of protons is given by $n_p=\rho N_A\left\langle Z/A\right\rangle $ and the density 
of neutrons by $n_n=\rho N_A(1-\left\langle Z/A\right\rangle )$. $N_A=6.022\times 10^{23}$\usk mol$^{-1}$ is 
the Avogadro constant, $\rho$ is the density of the material and $\left\langle Z/A\right\rangle $ is the 
average ratio of the proton number and the mass number of the material as given in Table~\ref{materials}. Due 
to the small neutrino--nucleon cross-section the attenuation term that accounts for the absorption of part of 
the signal and background neutrino fluxes during the passage of the Earth is negligible in the considered 
energy range. However, since the neutrino--nucleon cross-section rises with increasing neutrino energy, this 
effect becomes non-negligible at neutrino energies above 10\usk TeV. 

The neutrino effective area $A^{\text{eff}}_{\nu_{\mu}}$ is defined as the ratio of the rate of reconstructed 
events and the incident neutrino flux. It is calculated using Monte Carlo methods and incorporates the 
attenuation of the neutrino flux during the passage of the Earth, the neutrino--nucleon cross-section, the 
range of the generated muon and the reconstruction and selection efficiencies. This effective area is usually 
provided by the experimental collaborations. The energy dependence of the neutrino effective area comes mainly 
from the energy dependence of the cross-section (roughly $\propto E_{\nu}$) and the increase of the muon range. 
Notice that the muon effective area $A^{\text{eff}}_{\mu}$, on the other hand, is defined as the ratio of the 
rate of reconstructed events and the incident muon flux. This area incorporates only the geometry of the 
detector and the detection efficiency. It is roughly equal to the geometrical area but might have a slight 
energy dependence. 

For the calculation of the spectrum of muon neutrino induced muons at the detector position we have to take 
into account the shift to lower energies due to the energy loss during muon propagation through 
matter~\cite{Erkoca:2009by}: 
\begin{align}
 \frac{d\phi_{\mu}}{dE_{\mu}}= &\int d\Omega\int_{E_{\mu}}^{\infty} dE_{\nu_{\mu}}\!\!\!\!\!\!\!\!\int_0^{R_{\mu}(E_{\nu_{\mu}},E_{\mu})}\!\!\!\!\!\!\!\!\!dr\,e^{\beta\varrho r}\, \frac{dJ(E_{\nu_{\mu}},\theta,\phi)}{dE_{\nu_{\mu}}}\left[ \frac{d\sigma_{\text{CC}}^{\nu p}(E_{\nu_{\mu}},E_{\mu}^0)}{dE_{\mu}^0}\,n_p+(p\rightarrow n)\right] _{E_{\mu}^0=E_{\mu}^0(E_{\mu})} \nonumber\\ 
 &+(\nu\rightarrow\bar{\nu})\,,
 \label{through-spec}
\end{align}
where we neglected the attenuation term. In this expression the initial muon energy enters as an explicit 
function of the final muon energy as given by equation~(\ref{muonenergy}). 

\begin{figure}
 \centering
 \includegraphics[scale=0.9,bb=99 56 354 298,clip]{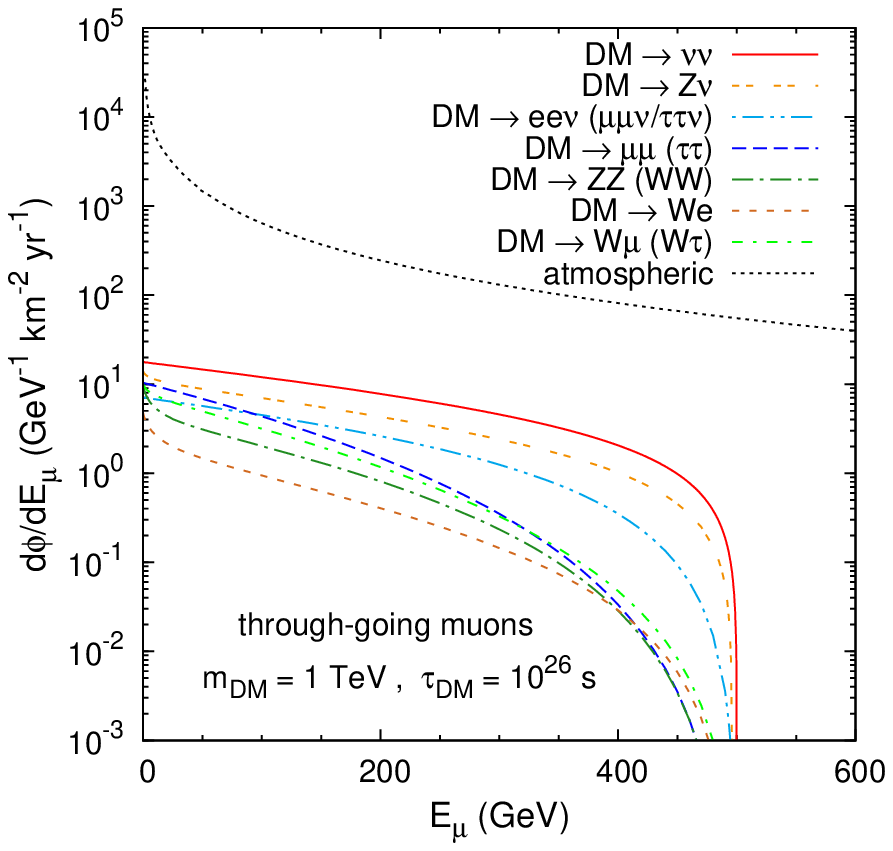}
 \includegraphics[scale=0.9,bb=117 56 356 298,clip]{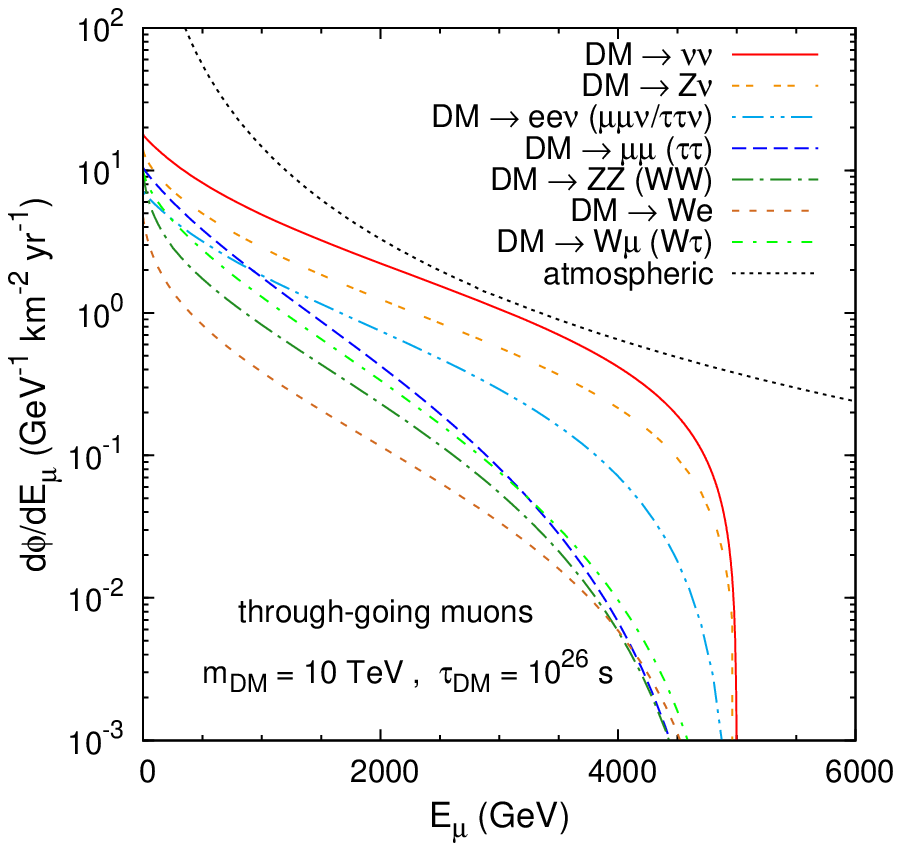}
 \caption{Muon fluxes for the different decay channels of a dark matter candidate compared to the atmospheric 
 background for upward through-going muons in standard rock. The flux is computed for a dark matter mass of 
 1\usk TeV (\textit{left}) or 10\usk TeV (\textit{right}) and a lifetime of $10^{26}$\usk s, for the neutrino 
 spectra in Figures~\ref{Spectra-plot} and~\ref{Spectra-plot2}. In this case the muons lose energy on their 
 way to the detector, smoothing out the spectral edges.}
 \label{SK-Mu-spectra}
\end{figure}
Using equation~(\ref{through-spec}) we calculate the flux of through-going muons induced by neutrinos from 
various dark matter decay channels and show the results in Figure~\ref{SK-Mu-spectra} for the case of a 
detector surrounded by standard rock. However, the result is also applicable for the case of detectors 
surrounded by water or ice since the dependence on the density cancels in equation~(\ref{through-spec}) and 
the muon energy loss parameters are roughly similar for the different materials (\textit{cf.} 
Table~\ref{materials}). Since there is no possibility to veto for the overwhelming background of atmospheric 
muons, only upgoing events and therefore a solid angle of $2\pi$ can be used for the analysis. We see that the 
deep inelastic scattering transforms the monochromatic neutrino lines into a continuous muon spectrum. In 
addition, the energy loss in the muon propagation smooths out all the spectra making the edge corresponding 
to half the dark matter particle mass less clear. Still the spectrum for a line signal remains steeper than 
the others at the endpoint.

\subsubsection{Contained Muons}

These events are similar to through-going muons but in this case the neutrino--nucleon interaction takes 
place inside the instrumented volume. If the muon track ends inside the detector the events are called 
contained. If the muon track leaves the detector one speaks of a partially contained event. The rate of muon 
neutrino induced (partially) contained track-like events per unit detector volume is given by 
\begin{equation}
  \frac{dN}{dE_{\mu}\,dVdt} =\int d\Omega\int_{E_{\mu}}^{\infty} dE_{\nu_{\mu}}\, \frac{dJ_{\nu_{\mu}}(E_{\nu_{\mu}},\theta,\phi)}{dE_{\nu_{\mu}}}\left[ \frac{d\sigma_{\text{CC}}^{\nu p}(E_{\nu_{\mu}},E_{\mu})}{dE_{\mu}}\,n_p+(p\rightarrow n)\right] +(\nu\rightarrow\bar{\nu})\,,
 \label{contained-rate}
\end{equation}
where we also neglected the attenuation term. In this case also the hadronic cascade is contained in the 
detector volume and therefore, by measuring the energy of the muon as well as of the hadronic cascade, it is 
in principle possible to reconstruct the total energy of the incident muon neutrino. In this case, however, 
one has to rely on the detection also of the hadronic cascade which, as we will discuss later, seems to be 
challenging. 
\begin{figure}
 \centering
 \includegraphics[scale=0.9,bb=99 56 354 294,clip]{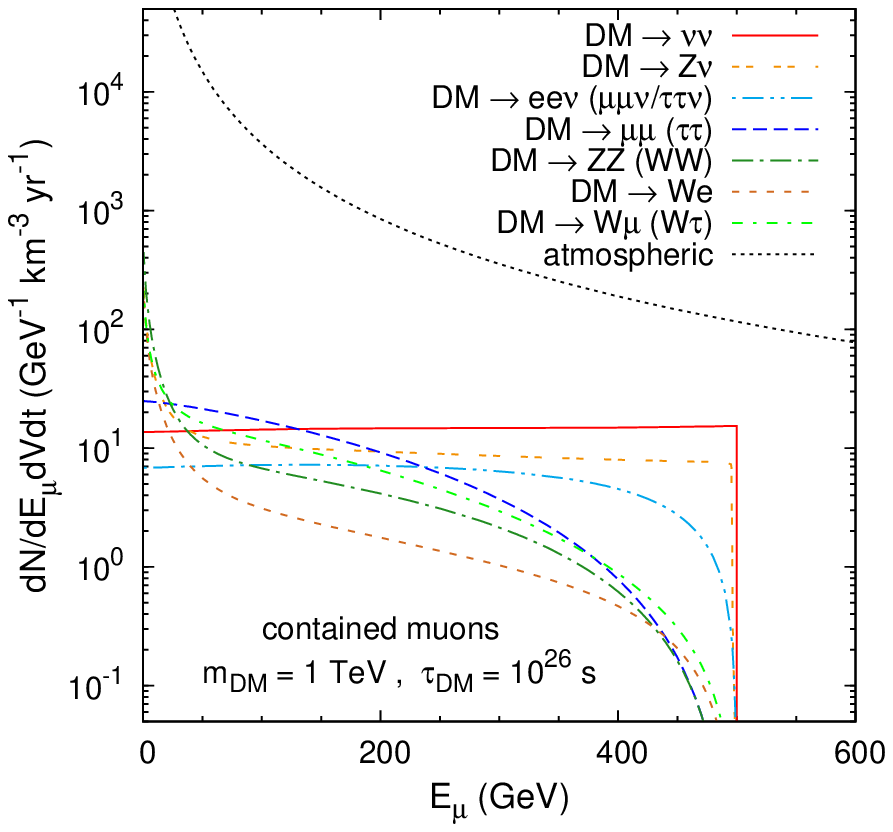}
 \includegraphics[scale=0.9,bb=117 56 356 294,clip]{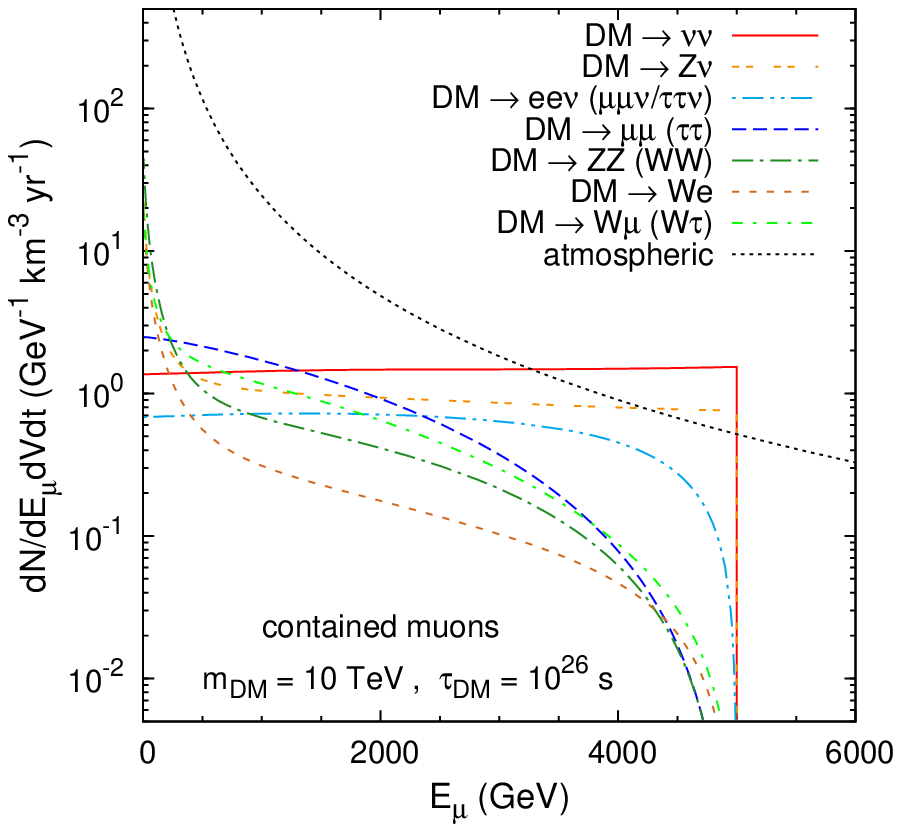}
 \caption{Spectra of contained muons for the different decay channels of a dark matter candidate compared to 
 the atmospheric background. The event rate per km$^3$ of detector volume (filled with ice) is computed for a 
 dark matter mass of 1\usk TeV (\textit{left}) or 10\usk TeV (\textit{right}) and a lifetime of $10^{26}$\usk s, 
 for the neutrino spectra in Figures~\ref{Spectra-plot} and \ref{Spectra-plot2}. The line signal is changed 
 into a muon continuum due to deep inelastic scattering, but it retains a hard edge at half the decaying 
 particle mass. Also the other spectra are softer than the original neutrino ones, with the continuum neutrino 
 spectra producing a rise at low energy, unfortunately well below the background.}
 \label{Mu-cont-spectra}
\end{figure}

The effective volume of the detector for contained events corresponds roughly to the geometrical volume (apart 
from boundary effects and reconstruction efficiency) and it is not enhanced by the muon range, which as we 
have seen, grows as $E_\nu$. Therefore, the statistics for contained events is much lower than for 
through-going events at large energies. For instance in the case of Super-Kamiokande the event rate above 
roughly 10\usk GeV is dominated by through-going muons. On the other hand, in the energy range of interest for 
dark matter searches the muon range is of the order of one kilometer and therefore the expected rate of 
contained muons is comparable to the rate of through-going muons in detectors of cubic kilometer size. Thus, these 
contained events might be equally important for dark matter searches at the new generation of neutrino 
telescopes. 

In addition, for downgoing contained muon events there is the interesting possibility to reduce the background 
of atmospheric muon neutrinos by the detection of a coincident muon that was produced in the same parent meson 
decay~\cite{Schonert:2008is}. This strategy could be used to increase the signal-to-background ratio for this 
channel, especially at large energies. However, we will not discuss this strategy quantitatively in this work. 

In Figure~\ref{Mu-cont-spectra} we show the muon spectra for contained events calculated using 
equation~(\ref{contained-rate}) for the case of a detector volume filled with ice. The result for a volume of 
water can easily be obtained rescaling the rate with the slightly different density. In this case there is no 
smoothing due to muon energy loss as in the case of through-going muons and the edges of the spectra are 
clearer, in particular for the case of a two-body decay. 

Here we only discussed the case where only the muon is measured since this is what can be done by the 
experiments at the moment. If the hadronic shower is also measured the combined reconstructed spectra would be 
as in Figures~\ref{Spectra-plot} and \ref{Spectra-plot2}. This is similar to the case of electron and tau 
neutrinos that is discussed in the next section. However, as will be discussed there, that channel offers a 
better signal-to-background ratio and a better energy resolution and will therefore be of more interest once 
the showers can be measured and used for analyses.

\subsection{Electron and Tau Neutrinos}
\label{showers}

The charged-current deep inelastic scattering of an electron neutrino off a nucleus produces a hadronic shower 
and an electron that immediately causes an electromagnetic shower. The charged-current deep inelastic 
scattering of a tau neutrino off a nucleus produces a hadronic shower and a tau lepton. Due to the short 
lifetime of the tau lepton ($c\tau_\tau=87.11\usk\mu$m), at these energies it decays almost instantly and 
produces another shower at the interaction point. Thus, at energies below many TeV, detectors like IceCube 
cannot distinguish electron neutrino from tau neutrino events since both types produce similar showers in the 
detector~\cite{Cowen:2007ny}. In these cases, however, the whole neutrino energy is deposited in the detector
and therefore it may be possible in principle to reconstruct better the initial neutrino spectrum. On the 
other hand, the analysis for cascade-like events is much more difficult than the analysis for muon tracks. No 
cascade events from atmospheric neutrinos have been identified yet and there are only first studies on this 
topic \textit{e.g.} by the IceCube collaboration~\cite{D'Agostino:2009sj}. For this reason there is no effective 
area for this type of events available yet and therefore it is difficult to estimate realistically the 
sensitivity in shower events.

Shower-like events are also characteristic of the neutrino--nucleon neutral-current interaction and for this 
reason probably only a combined analysis of neutral-current interactions for all neutrino flavours and 
charged-current interactions for tau and electron neutrinos will be feasible. In this case the total rate of 
neutrino-induced shower-like events is given by 
\begin{align}
 \frac{dN}{dE_{\text{shower}}\,dVdt}= &\int d\Omega\,\Bigg\lbrace \sum_{\ell=e,\tau}\left( \frac{dJ_{\nu_\ell}(E_{\nu_\ell},\theta,\phi)}{dE_{\nu_\ell}}\left[ \sigma_{\text{CC}}^{\nu p}(E_{\nu_\ell})\,n_p+(p\rightarrow n)\right] \right) _{E_{\nu_\ell}=E_{\text{shower}}} \nonumber\\
 &+\sum_{\ell=e,\mu,\tau}\,\int_{E_{\text{shower}}}^{\infty} \!\!\!\!dE_{\nu_\ell}\, \frac{dJ_{\nu_\ell}(E_{\nu_\ell},\theta,\phi)}{dE_{\nu_\ell}}\left[ \frac{d\sigma_{\text{NC}}^{\nu p}(E_{\nu_\ell},E_{\text{shower}})}{dE_{\text{shower}}}\,n_p+(p\rightarrow n)\right] \Bigg\rbrace \nonumber\\  &+(\nu\rightarrow\bar{\nu})\,.
 \label{cascade-rate}
\end{align}
\begin{figure}
 \centering
 \includegraphics[scale=0.9,bb=101 57 352 294,clip]{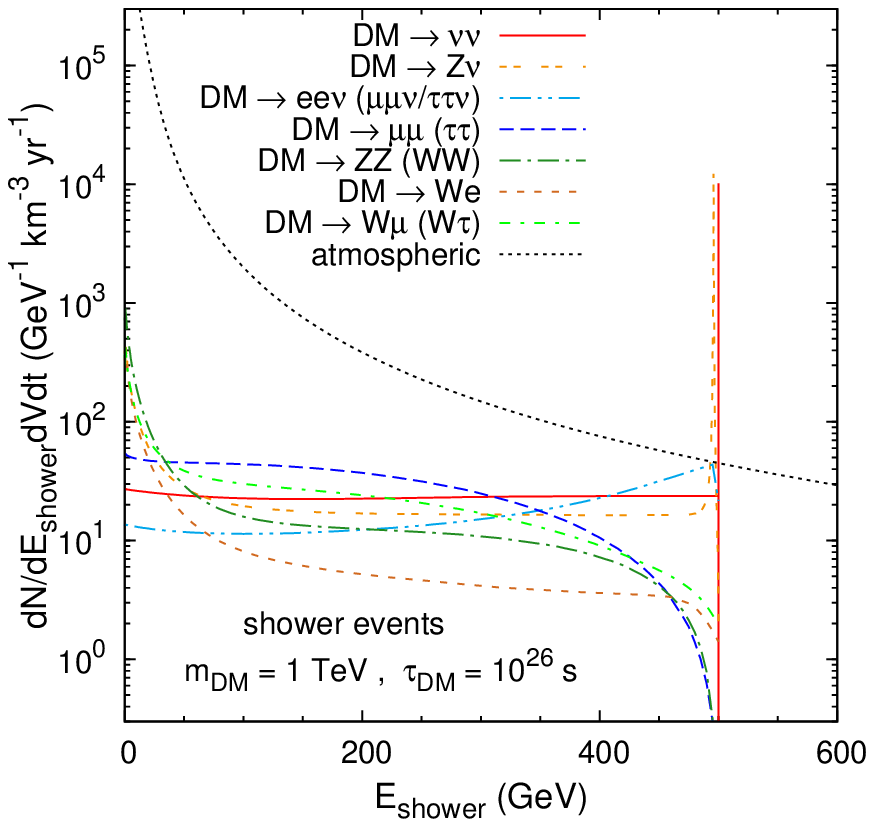}
 \includegraphics[scale=0.9,bb=117 57 356 294,clip]{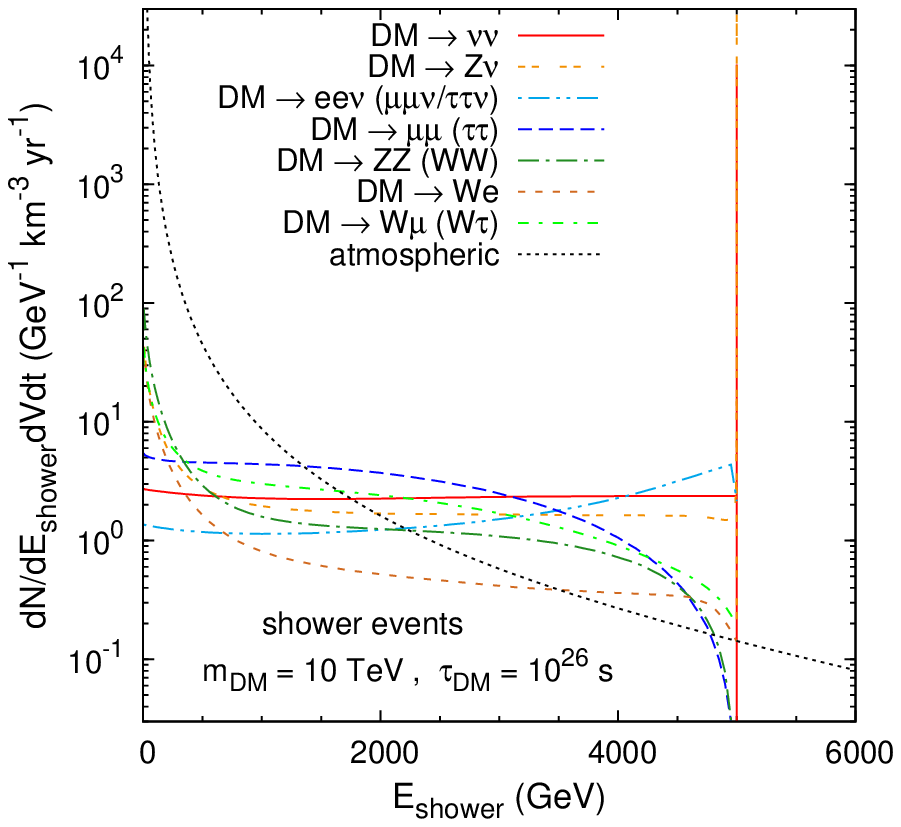}
 \caption{Spectra of cascade-like events for the different decay channels of a dark matter candidate compared 
 to the atmospheric background. The event rate per km$^3$ of detector volume (filled with ice) is computed 
 for a dark matter mass of 1\usk TeV (\textit{left}) or 10\usk TeV (\textit{right}) and a lifetime of 
 $10^{26}$\usk s, for the neutrino spectra in Figures~\ref{Spectra-plot} and~\ref{Spectra-plot2}.}
 \label{IC-Shower-spectra}
\end{figure}

We give in Figure~\ref{IC-Shower-spectra} the signal and atmospheric background spectra calculated from 
equation~(\ref{cascade-rate}) for the case of a detector volume filled with ice. Note that in this case the 
muon neutrinos contribute only via neutral-current interactions which are weaker by a factor of about three 
compared to charged-current interactions (\textit{cf.} equation~(\ref{totalcross})). Still, since the 
atmospheric muon neutrino flux is a factor of 20 larger than the electron neutrino flux at TeV energies, the 
atmospheric muon neutrinos provide the dominant background. At the same time the signal is increased by 
roughly a factor of three. This is because, due to neutrino oscillations, the signal is roughly equal in all 
neutrino flavours and, therefore, the signal rate from the charged-current interactions of electron and tau 
neutrinos is the same as for the muon neutrinos. In addition, the combined neutral-current signal of all 
flavours contributes at the same level as the charged-current signal of one flavour. 
In summary, cascade-like events will offer a signal-to-background ratio that is roughly one order of magnitude 
larger than in the muon case and, therefore, they appear to be a very promising channel, if they are measured. 
We see also that in this case, assuming that the total shower energy can be reconstructed, the line-feature is 
preserved and clearly visible.

\section{Rates and Bounds}
\label{Bounds}

\subsection{Super-Kamiokande}

\begin{figure}
 \centering
 \includegraphics[scale=.91,bb=102 58 352 293,clip]{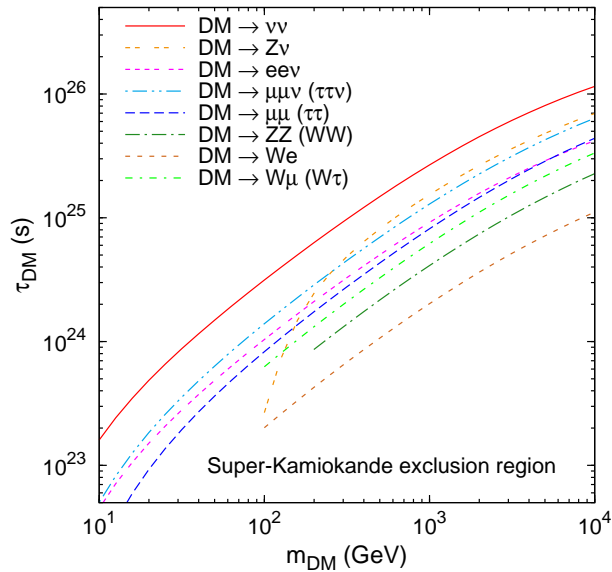}
 \caption{90\usk\% C.L. exclusion region in the lifetime vs. mass plane for a decaying dark matter candidate 
 from the non-observation of an excess in the Super-K data. The bound is stronger for a line signal, since 
 there the spectrum is harder, resulting in a larger muon flux due to the increasing neutrino--nucleon 
 cross-section and muon range. For the channels that contain $Z^0$ or $W^\pm$ bosons in the final state the 
 exclusion range is cut at the threshold for their production.}
\label{SK-limit}
\end{figure}

Super-Kamiokande is a 50\usk kt water Cherenkov detector. The fiducial mass is 22.5\usk kt and the muon 
effective area is 1200\usk m$^2$ (with a slight zenith angle dependence due to the cylindrical shape of the 
detector). It is identical to the geometrical area since the reconstruction and selection efficiencies are 
virtually 100\usk\%. Super-Kamiokande has been looking for a neutrino signal, mostly from dark matter 
annihilation in the centre of the Sun, the centre of the Earth and in the Galactic centre. No excess has been 
found so far, and this can also be used to put a constraint on the decaying dark matter case. We compare the 
flux of upward through-going muons from dark matter decay (integrated over energies above the threshold at 
1.6\usk GeV) with the 90\usk\% C.L. flux limit of excess neutrino-induced upward through-going muons provided 
by the Super-Kamiokande collaboration for the Galactic centre (the limits from the Sun and Earth flux are 
weaker)~\cite{Desai:2004pq}. As discussed in section~\ref{direction}, the strongest bounds are obtained for 
the largest field of view. The exclusion region in the decaying dark matter parameter space derived from the 
limit in the $30^\circ$ half-angle cone around the Galactic centre is given in Figure~\ref{SK-limit}. 

The bounds obtained here become stronger for larger masses although the neutrino flux is proportional to 
$1/m_\text{DM}$ for constant lifetime. This is due to the increasing neutrino--nucleon cross-section and the 
increasing muon range. The bounds are stronger for the two-body decay signal compared to the other cases since 
there the signal is concentrated at the end of the spectrum and benefits from the larger neutrino--nucleon 
cross-section and muon range. Note that these present bounds do not have sensitivity to the parameter region 
preferred by the PAMELA excess yet, which corresponds to a lifetime of the order of $10^{26}$\usk s and masses 
larger than 200\usk GeV.

\subsection{Rates and Bounds for Present and Future Experiments}

Assuming decaying dark matter with a lifetime of $10^{26}$\usk s, we can now compute the expected signal rates 
for present and future experiments. These results can be easily generalised to arbitrary lifetimes, by 
recalling that the flux is proportional to $1/\tau_\text{DM}$. We give the rates for some typical detectors of 
different sizes, \textit{i.e.} Super-Kamiokande, ANTARES/AMANDA and IceCube. The results for Super-K can be 
easily scaled up to the Hyper-Kamiokande/UNO size by multiplying by a factor 10 or 20 (for a Hyper-K mass of 
500\usk kt and Hyper-K/UNO mass of 1\usk Mt, respectively). The result for KM3NeT will be very similar to that 
expected for IceCube.

We would like to stress here that Super-K is still taking data, and that the full ANTARES detector was 
completed in summer 2008 and is also operational. The AMANDA detector was decommissioned in summer 2009, 
but has since been substituted by the partial IceCube detector, which already had 59 strings deployed in 
the ice in early 2009. The other experiments are still in the planning phase: KM3NeT is a proposed 
cubic-kilometer sized underwater neutrino telescope in the Mediterranean Sea, which will probably have an 
effective volume comparable to IceCube, but will be able to look at the Galactic centre, while the proposed 
Hyper-Kamiokande and Underwater Neutrino Observatory (UNO) are water Cherenkov detectors similar to 
Super-Kamiokande but of megaton scale.

For the case of IceCube we also take into account the DeepCore subdetector, which is currently under 
construction. It is designed to lower the energy threshold of the experiment to roughly 30\usk GeV and to increase 
the sensitivity at low energies. This detector consists of six additional strings with less spacing between 
the digital optical modules compared to IceCube. One of these strings has already been deployed in early 2009. 
Commissioning of the full detector is planned for early 2010, while IceCube itself will only be completed in 
early 2011. The combination of IceCube and DeepCore can use the outer layers of IceCube as a veto to 
atmospheric muons and therefore has a $4\pi$ sensitivity for fully and partially contained events, but a 
considerably smaller effective volume.

For the calculation of rates of upward through-going muon events we use the neutrino effective areas for 
AMANDA, ANTARES and the IceCube 80 strings configuration from~\cite{Montaruli:2009kv} and integrate over 
the muon spectrum. For the combined IceCube + DeepCore detector we amend the effective area in the 
low-energy range using the neutrino effective area given in~\cite{Wiebusch:2009jf}. In the case of 
Super-Kamiokande we calculate the rate using equation~(\ref{through-rate}) with standard rock as surrounding 
material, a muon effective area of 1200\usk m$^2$ and a threshold muon energy of 1.6\usk GeV. 

\begin{table}
 \centering
 \begin{tabular}{lccccc}
  \hline
  decay channel & Super-K & AMANDA & ANTARES & IceCube & IC+DeepCore \\
  \hline
  atmospheric $\nu_\mu$ & $4.3\times10^2$ & $1.5\times10^3$ & $1.8\times10^3$ & $3.0\times10^5$ & $3.5\times10^5$ \\
  \hline
  $DM\rightarrow\nu\bar{\nu}$ & $1.4\times10^0$ & $5.0\times10^0$ & $6.4\times10^0$ & $1.4\times10^3$ & $1.6\times10^3$ \\
  $DM\rightarrow\mu^+\mu^-$ & $4.1\times10^{-1}$ & $6.3\times10^{-1}$ & $1.0\times10^{0}$ & $2.7\times10^2$ & $3.5\times10^2$ \\
  $DM\rightarrow\tau^+\tau^-$ & $4.1\times10^{-1}$ & $6.0\times10^{-1}$ & $9.6\times10^{-1}$ & $2.6\times10^2$ & $3.5\times10^2$ \\
  $DM\rightarrow Z^0Z^0$ & $2.4\times10^{-1}$ & $3.6\times10^{-1}$ & $5.4\times10^{-1}$ & $1.4\times10^2$ & $1.8\times10^2$ \\
  $DM\rightarrow W^+W^-$ & $1.8\times10^{-1}$ & $3.0\times10^{-1}$ & $4.5\times10^{-1}$ & $1.2\times10^2$ & $1.5\times10^2$ \\  
  $DM\rightarrow Z^0\nu$ & $7.1\times10^{-1}$ & $2.1\times10^0$ & $2.8\times10^0$ & $6.4\times10^2$ & $7.3\times10^2$ \\
  $DM\rightarrow e^+e^-\nu$ & $4.8\times10^{-1}$ & $1.4\times10^0$ & $1.9\times10^0$ & $4.4\times10^2$ & $5.0\times10^2$ \\
  $DM\rightarrow\mu^+\mu^-\nu$ & $6.7\times10^{-1}$ & $1.6\times10^0$ & $2.2\times10^0$ & $5.4\times10^2$ & $6.4\times10^2$ \\
  $DM\rightarrow\tau^+\tau^-\nu$ & $6.8\times10^{-1}$ & $1.6\times10^0$ & $2.2\times10^0$ & $5.3\times10^2$ & $6.4\times10^2$ \\
  $DM\rightarrow W^\pm e^\mp$ & $1.0\times10^{-1}$ & $2.0\times10^{-1}$ & $2.8\times10^{-1}$ & $7.1\times10^1$ & $8.9\times10^1$ \\
  $DM\rightarrow W^\pm\mu^\mp$ & $3.1\times10^{-1}$ & $5.2\times10^{-1}$ & $8.0\times10^{-1}$ & $2.1\times10^2$ & $2.7\times10^2$ \\
  $DM\rightarrow W^\pm\tau^\mp$ & $2.9\times10^{-1}$ & $4.4\times10^{-1}$ & $7.0\times10^{-1}$ & $1.9\times10^2$ & $2.4\times10^2$ \\
  \hline
 \end{tabular}
 \caption{Number of upward through-going muon events per year from the atmospheric neutrino background and 
 different dark matter decay channels at several neutrino experiments. The signals are given for a dark matter 
 lifetime of $10^{26}$\usk s and a dark matter mass of 300\usk GeV.}
 \label{rates1}
\end{table}

\begin{table}
 \centering
 \begin{tabular}{lccccc}
  \hline
  decay channel & Super-K & AMANDA & ANTARES & IceCube & IC+DeepCore \\
  \hline
  atmospheric $\nu_\mu$ & $4.3\times10^2$ & $1.5\times10^3$ & $1.8\times10^3$ & $3.0\times10^5$ & $3.5\times10^5$ \\
  \hline
  $DM\rightarrow\nu\bar{\nu}$ & $4.1\times10^0$ & $2.4\times10^1$ & $2.8\times10^1$ & $4.6\times10^3$ & $4.6\times10^3$ \\
  $DM\rightarrow\mu^+\mu^-$ & $1.3\times10^0$ & $6.0\times10^0$ & $7.2\times10^0$ & $1.4\times10^3$ & $1.4\times10^3$ \\
  $DM\rightarrow\tau^+\tau^-$ & $1.3\times10^0$ & $5.8\times10^0$ & $7.0\times10^0$ & $1.3\times10^3$ & $1.4\times10^3$ \\
  $DM\rightarrow Z^0Z^0$ & $7.3\times10^{-1}$ & $2.9\times10^0$ & $3.5\times10^0$ & $6.6\times10^2$ & $6.8\times10^2$ \\
  $DM\rightarrow W^+W^-$ & $5.7\times10^{-1}$ & $2.5\times10^0$ & $3.1\times10^0$ & $5.8\times10^2$ & $6.0\times10^2$ \\
  $DM\rightarrow Z^0\nu$ & $2.4\times10^0$ & $1.3\times10^1$ & $1.6\times10^1$ & $2.6\times10^3$ & $2.6\times10^3$ \\
  $DM\rightarrow e^+e^-\nu$ & $1.4\times10^0$ & $7.8\times10^0$ & $9.3\times10^0$ & $1.6\times10^3$ & $1.6\times10^3$ \\
  $DM\rightarrow\mu^+\mu^-\nu$ & $2.0\times10^0$ & $1.0\times10^1$ & $1.2\times10^1$ & $2.2\times10^3$ & $2.2\times10^3$ \\
  $DM\rightarrow\tau^+\tau^-\nu$ & $2.0\times10^0$ & $1.0\times10^1$ & $1.2\times10^1$ & $2.2\times10^3$ & $2.2\times10^3$ \\
  $DM\rightarrow W^\pm e^\mp$ & $3.1\times10^{-1}$ & $1.4\times10^0$ & $1.7\times10^0$ & $3.2\times10^2$ & $3.3\times10^2$ \\
  $DM\rightarrow W^\pm\mu^\mp$ & $9.6\times10^{-1}$ & $4.5\times10^0$ & $5.4\times10^0$ & $1.0\times10^3$ & $1.0\times10^3$ \\
  $DM\rightarrow W^\pm\tau^\mp$ & $9.1\times10^{-1}$ & $4.0\times10^0$ & $4.9\times10^0$ & $9.3\times10^2$ & $9.8\times10^3$ \\
  \hline
  $DM\rightarrow\nu\bar{\nu}$ & $1.8\times10^1$ & $1.7\times10^2$ & $1.7\times10^2$ & $1.3\times10^4$ & $1.3\times10^4$ \\
  $DM\rightarrow\mu^+\mu^-$ & $6.8\times10^0$ & $5.5\times10^1$ & $6.0\times10^1$ & $5.5\times10^3$ & $5.5\times10^3$ \\
  $DM\rightarrow\tau^+\tau^-$ & $7.2\times10^0$ & $5.7\times10^1$ & $6.2\times10^1$ & $6.0\times10^3$ & $6.0\times10^3$ \\
  $DM\rightarrow Z^0Z^0$ & $3.8\times10^0$ & $2.7\times10^1$ & $3.0\times10^1$ & $2.9\times10^3$ & $2.9\times10^3$ \\
  $DM\rightarrow W^+W^-$ & $3.2\times10^0$ & $2.5\times10^1$ & $2.7\times10^1$ & $2.6\times10^3$ & $2.7\times10^3$ \\
  $DM\rightarrow Z^0\nu$ & $1.1\times10^1$ & $9.8\times10^1$ & $1.0\times10^2$ & $8.1\times10^3$ & $8.1\times10^3$ \\
  $DM\rightarrow e^+e^-\nu$ & $6.5\times10^0$ & $5.8\times10^1$ & $6.0\times10^1$ & $4.8\times10^3$ & $4.8\times10^3$ \\
  $DM\rightarrow\mu^+\mu^-\nu$ & $9.9\times10^0$ & $8.4\times10^1$ & $8.9\times10^1$ & $7.7\times10^3$ & $7.7\times10^3$ \\
  $DM\rightarrow\tau^+\tau^-\nu$ & $1.0\times10^1$ & $8.6\times10^1$ & $9.2\times10^1$ & $8.1\times10^3$ & $8.1\times10^3$ \\
  $DM\rightarrow W^\pm e^\mp$ & $1.7\times10^0$ & $1.3\times10^1$ & $1.5\times10^1$ & $1.4\times10^3$ & $1.4\times10^3$ \\
  $DM\rightarrow W^\pm\mu^\mp$ & $5.2\times10^0$ & $4.1\times10^1$ & $4.5\times10^1$ & $4.2\times10^3$ & $4.2\times10^3$ \\
  $DM\rightarrow W^\pm\tau^\mp$ & $5.1\times10^0$ & $4.0\times10^1$ & $4.3\times10^1$ & $4.2\times10^3$ & $4.2\times10^3$ \\
  \hline
 \end{tabular}
 \caption{Number of upward through-going muon events per year from the atmospheric neutrino background and 
 different dark matter decay channels at several neutrino experiments. The signals are given for a dark matter 
 lifetime of $10^{26}$\usk s and dark matter masses of 1\usk TeV \textit{(top)} and 10\usk TeV 
 \textit{(bottom)}.}
 \label{rates2}
\end{table}

We see from Tables~\ref{rates1} and \ref{rates2} that a sizable number of events is expected for a 
lifetime of $10^{26}$\usk s, especially for experiments of cubic kilometer scale, such as to become 
significant above the atmospheric background even for a dark matter particle mass of 300\usk GeV. 
Of course for larger masses the significance becomes greater due to the increasing signal rate. 
Note that here we did not make use of any spectral information. In that case larger dark matter 
masses would also benefit from the falling background. 

\begin{figure}
 \centering
 \includegraphics[scale=.91,bb=103 57 352 294,clip]{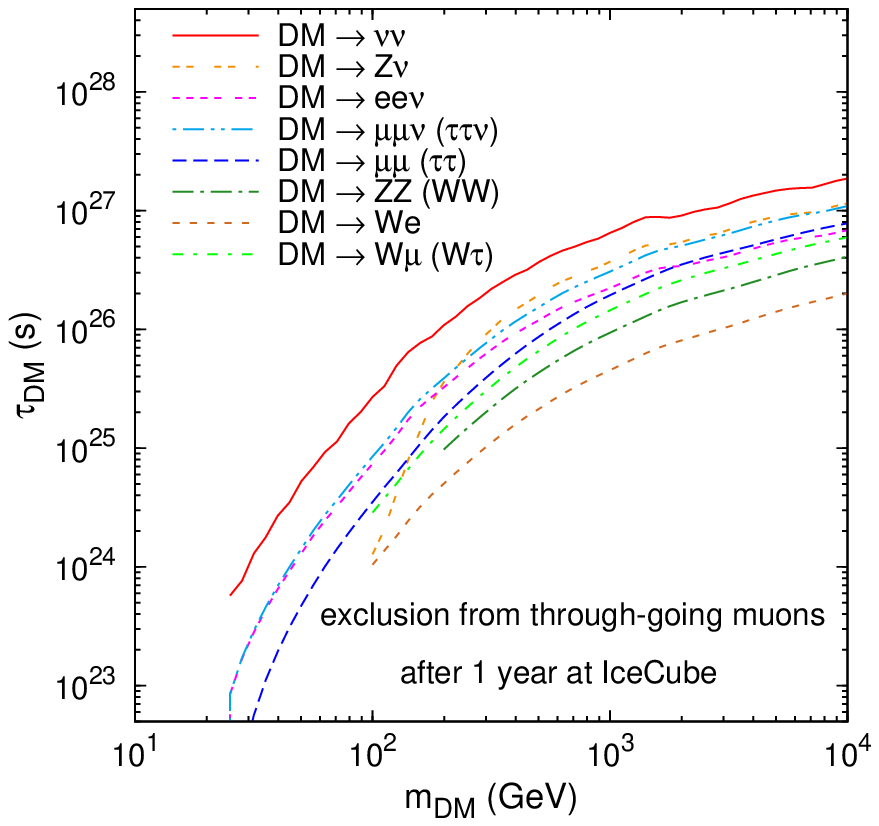}
 \includegraphics[scale=.91,bb=115 57 352 294,clip]{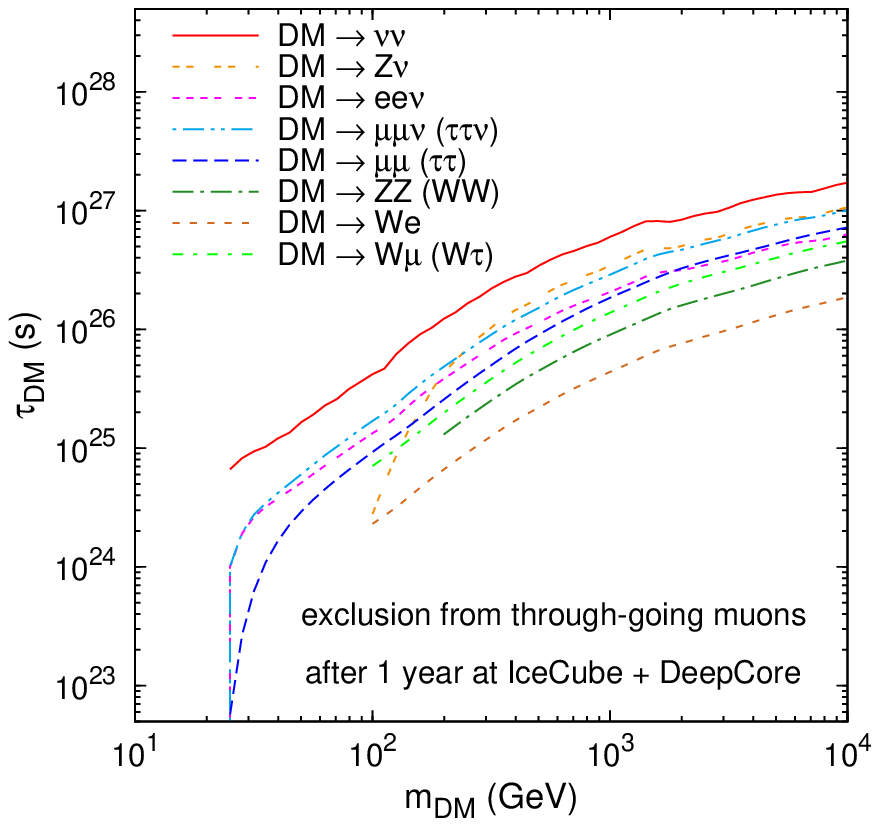}
 \caption{90\usk\% C.L. exclusion prospects in the lifetime vs. mass plane for a decaying dark matter 
 candidate from the non-observation of a significant excess in the total rate of neutrino induced upward through-going muons observed 
 at IceCube (\textit{left}) or IceCube + DeepCore (\textit{right}) in one year. Clearly visible is the enhanced sensitivity of the DeepCore extension in the 
 low-mass region.}
 \label{ICprospects}
\end{figure}

Requiring the combined number of signal and background events not to exceed the background above the 
90\usk\% C.L. (in the Gaussian approximation this corresponds to $\sigma=S/\sqrt{B}<1.28$), similar to 
the case of Super-K in Figure~\ref{SK-limit}, we can then give in Figure~\ref{ICprospects} a forecast of 
the exclusion region which may be obtained from kilometer-cubed experiments using one year of data. The 
larger statistics of the future experiments will improve the Super-K bounds by more than an order of 
magnitude and explore the region of lifetimes above $10^{25}$\usk s, for masses larger than 
200\usk GeV. Note that for ten years of data the lifetime limit will become stronger approximately by a 
factor of three. 
For lower masses the bounds will remain weaker, but in that parameter region a very
important role will be played by DeepCore, which will considerably improve the IceCube performance 
between 30--100\usk GeV masses as can be seen in the right panel of Figure~\ref{ICprospects}, and also 
by the megaton water detectors which are expected to strengthen the Super-K bounds by an order of magnitude 
down to masses of a few GeV. This low-mass region does not provide an explanation of the PAMELA excess
and is plagued by a stronger atmospheric background, but it is still remarkable that even there lifetimes 
larger than $10^{24}$\usk s will be probed in future experiments.

\subsection{Energy Resolution and Reconstructed Spectra}

Once a signal has been detected, the question arises if it will also be possible to reconstruct the 
neutrino spectra and extract some information on the dark matter decay channel. For this purpose one 
important factor is the energy resolution of the neutrino detectors. We will take here for reference the 
IceCube detector, for which the energy resolution is $\log_{10}(E_{\text{max}}/E_{\text{min}})=0.3$--0.4 
for track-like events and $\log_{10}(E_{\text{max}}/E_{\text{min}})=0.18$ for cascade-like 
events~\cite{Resconi:2008fe}. 

\begin{figure}
 \centering
 \includegraphics[scale=.92,bb=103 57 344 294,clip]{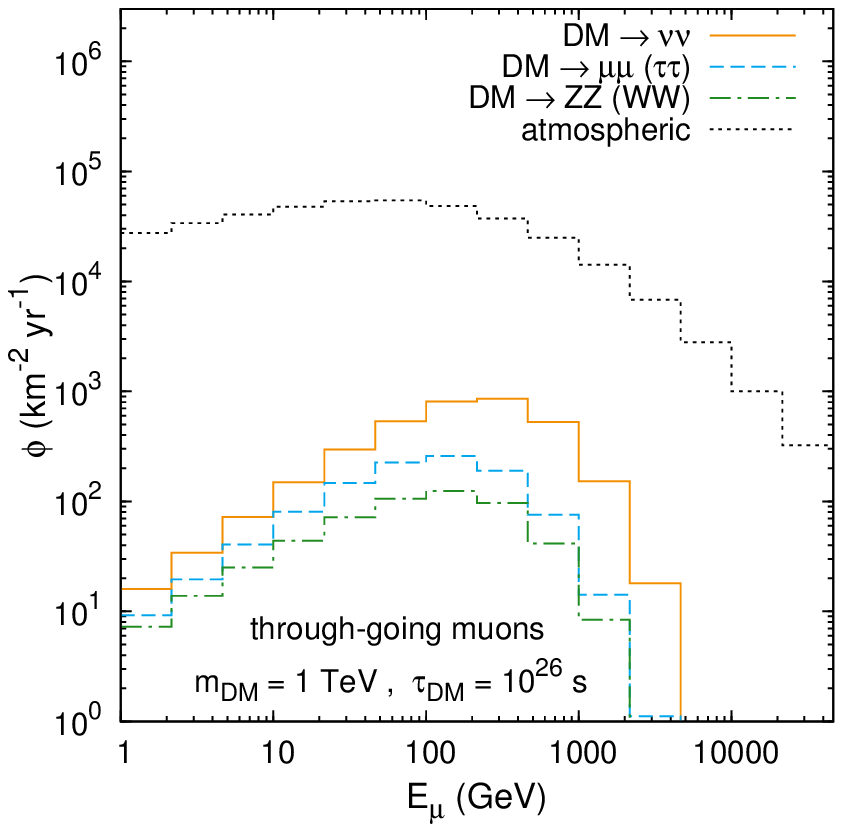}
 \includegraphics[scale=.92,bb=103 57 344 294,clip]{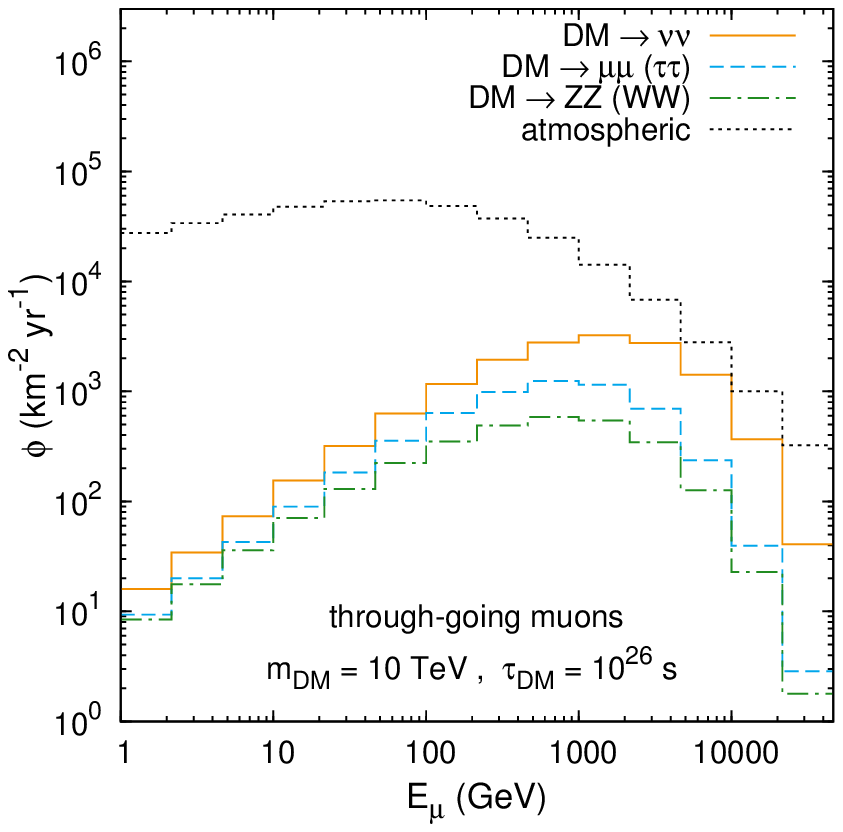}
 \includegraphics[scale=.92,bb=101 57 344 294,clip]{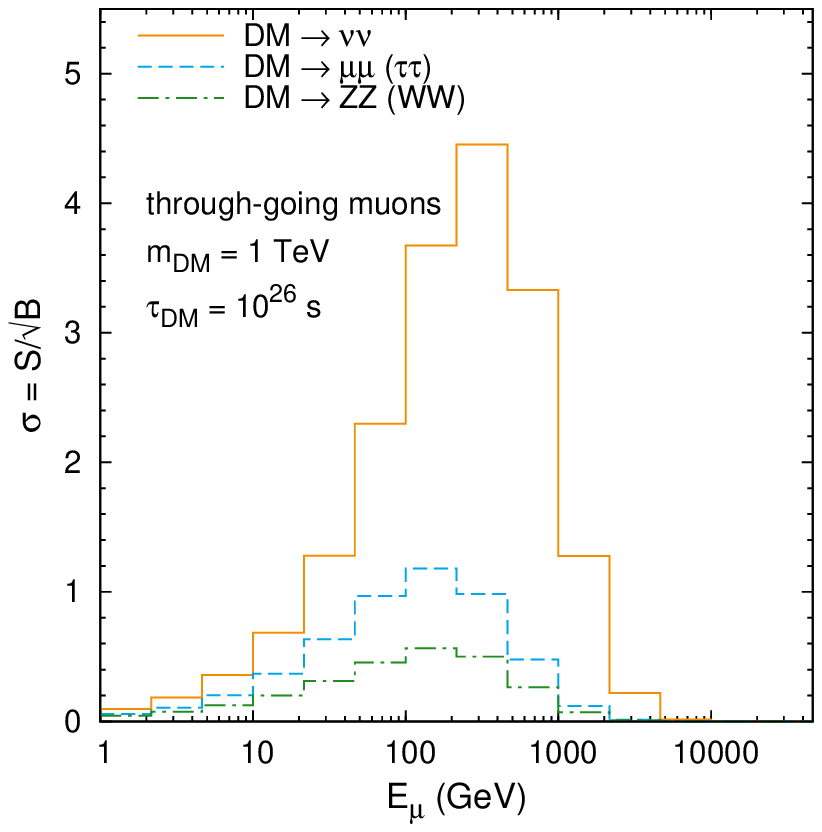}
 \includegraphics[scale=.92,bb=104 57 343 294,clip]{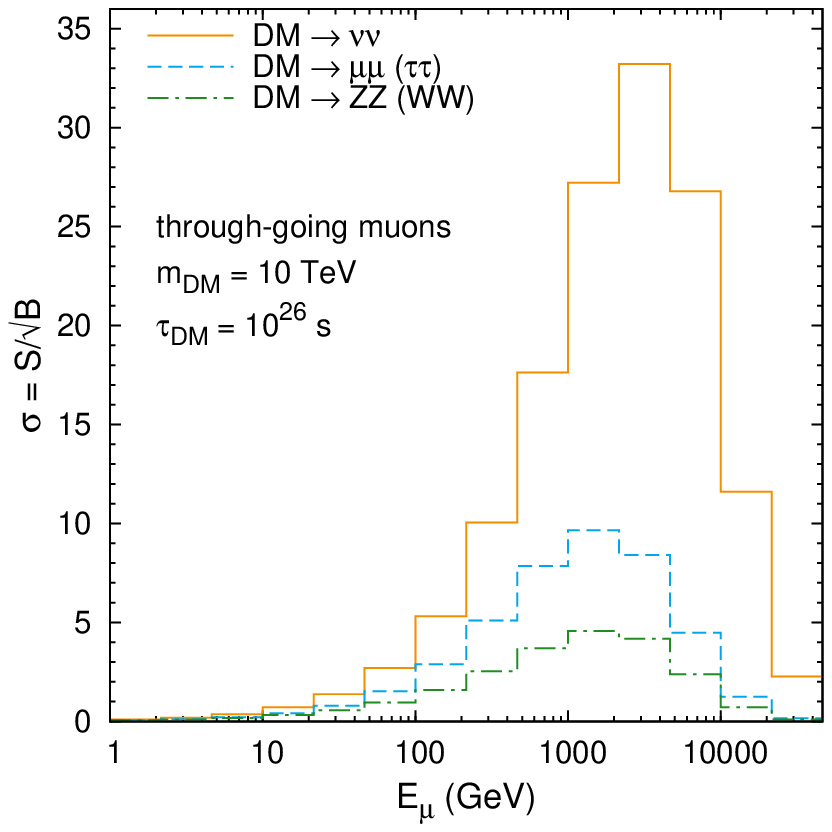}
 \caption{\textit{Top:} Muon fluxes for the different decay channels of a scalar dark matter candidate 
 compared to the atmospheric background for upward through-going muons. The flux is computed for a dark 
 matter mass of 1\usk TeV (\textit{left}) or 10\usk TeV (\textit{right}) and a lifetime of $10^{26}$\usk s 
 using an energy resolution of 0.3 in $\log_{10} E$ and three bins per decade. 
 \textit{Bottom:} Statistical significance of the signal of through-going muons shown above
 calculated for every single energy bin using one year of data with an effective area of 1\usk km$^2$.}
 \label{significance1a}
\end{figure}

\begin{figure}
 \centering
 \includegraphics[scale=.92,bb=103 57 344 294,clip]{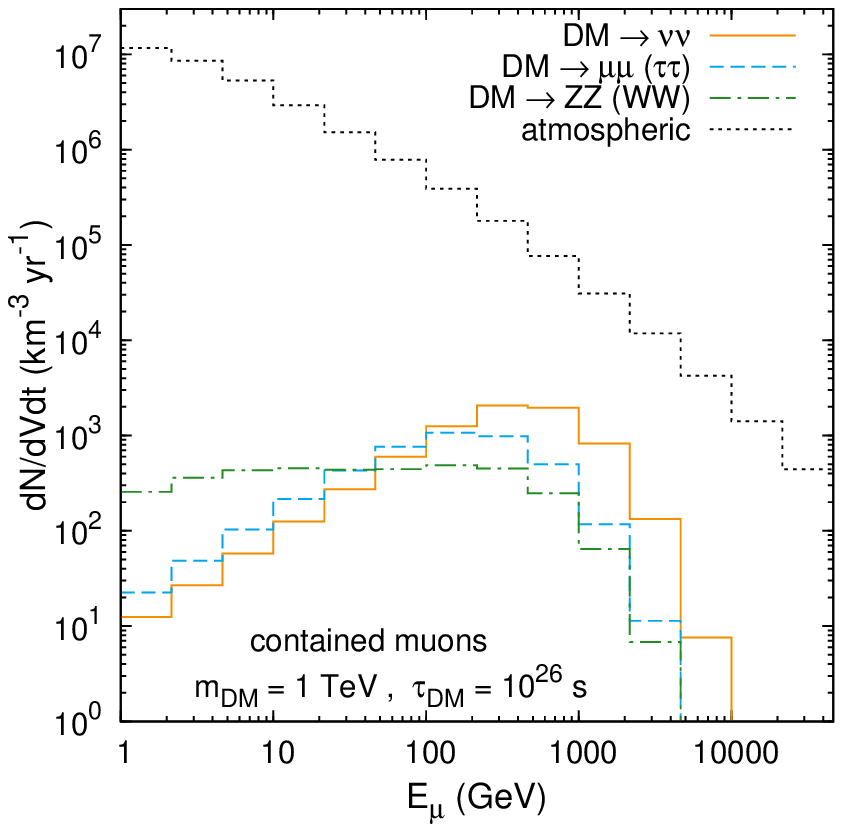}
 \includegraphics[scale=.92,bb=103 57 344 294,clip]{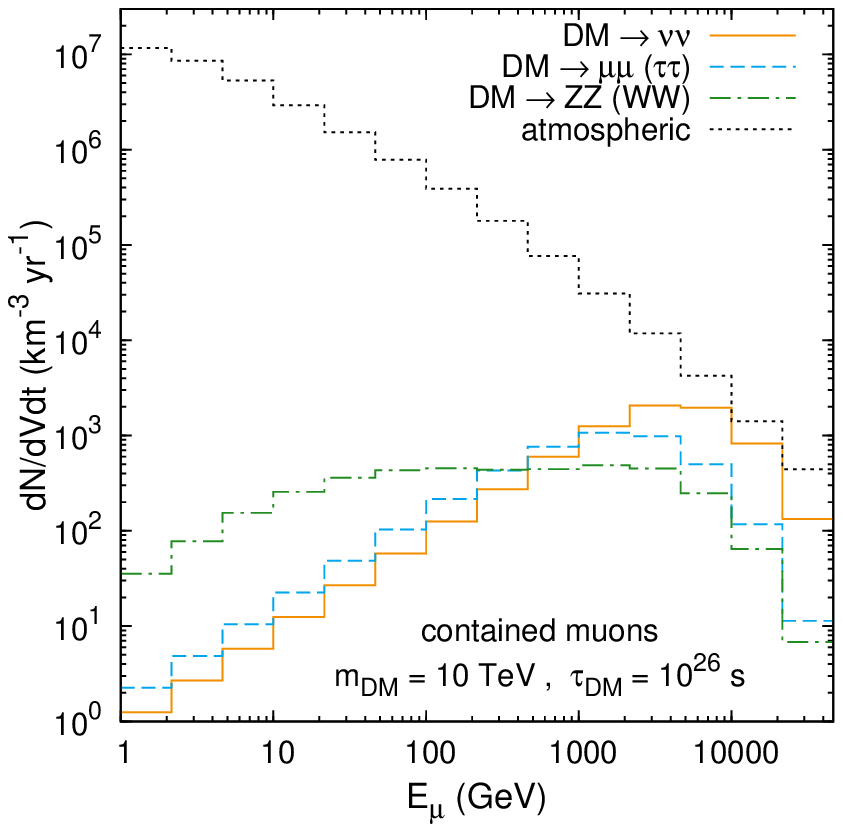}
 \includegraphics[scale=.92,bb=101 57 344 294,clip]{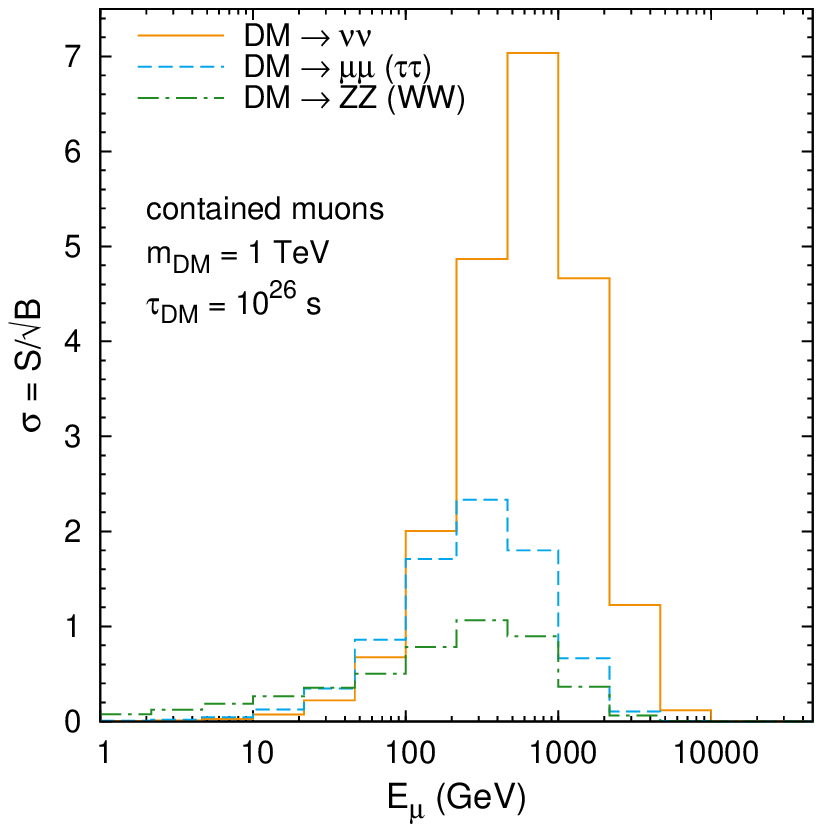}
 \includegraphics[scale=.92,bb=104 57 343 294,clip]{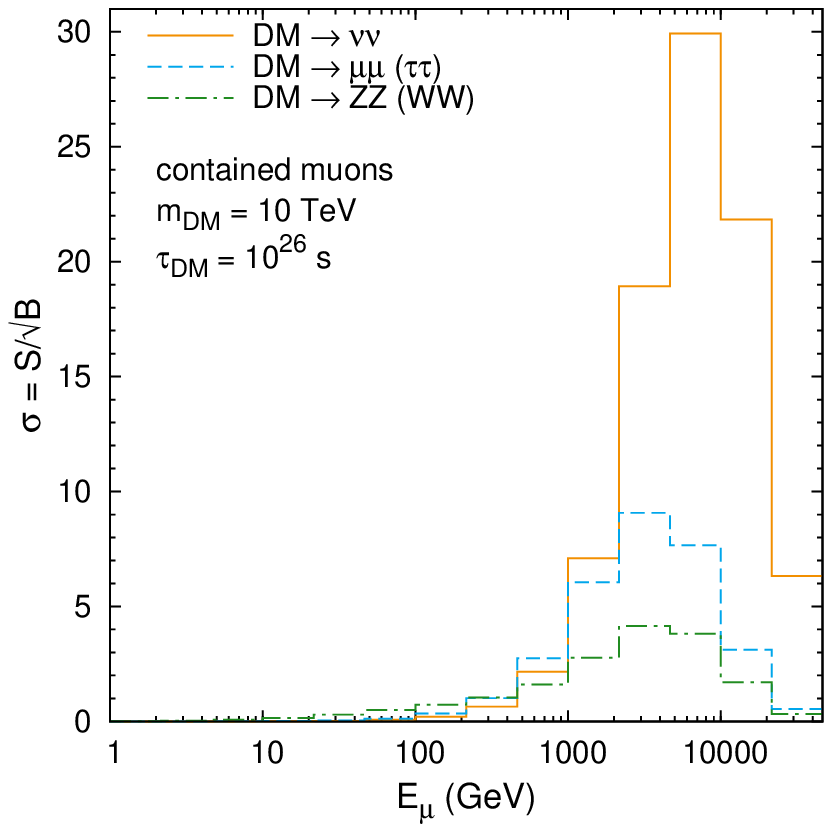}
 \caption{\textit{Top:} Muon rates per km$^3$ of detector volume for the different decay channels of a 
 scalar dark matter candidate compared to the atmospheric background for contained muons. The flux is computed 
 for a dark  matter mass of 1\usk TeV (\textit{left}) or 10\usk TeV (\textit{right}) and a lifetime of 
 $10^{26}$\usk s using an energy resolution of 0.3 in $\log_{10} E$ and three bins per decade. 
 \textit{Bottom:} Statistical significance of the contained muon signal shown above calculated 
 for every single energy bin using one year of data with an effective volume of 1\usk km$^3$.}
 \label{significance2a}
\end{figure}

\begin{figure}
 \centering
 \includegraphics[scale=.92,bb=103 57 344 294,clip]{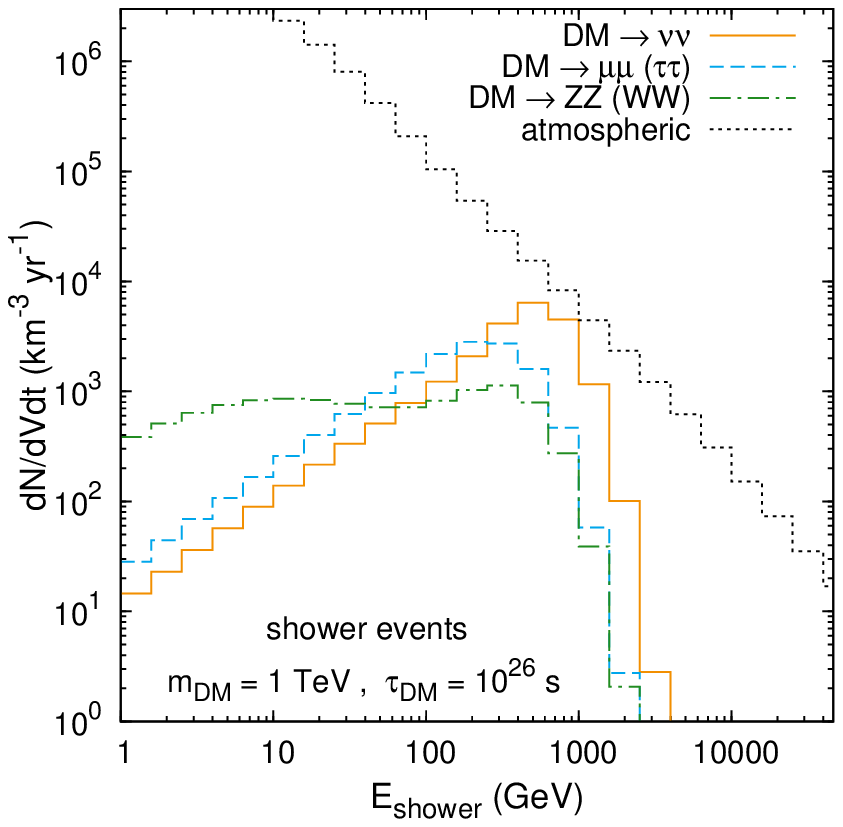}
 \includegraphics[scale=.92,bb=103 57 344 294,clip]{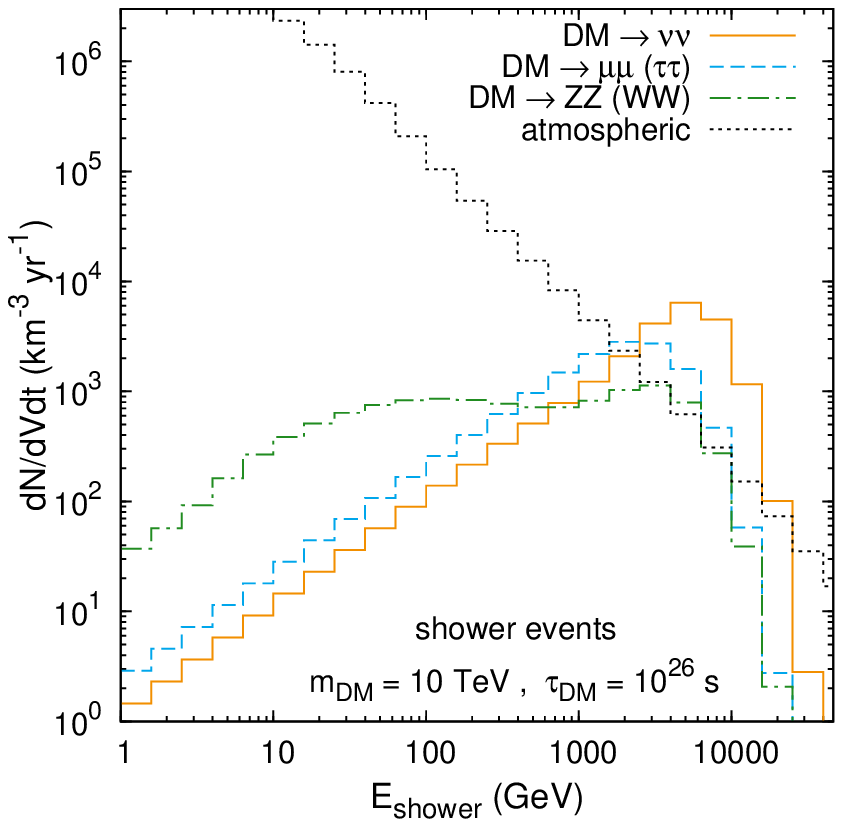}
 \includegraphics[scale=.92,bb=102 57 345 294,clip]{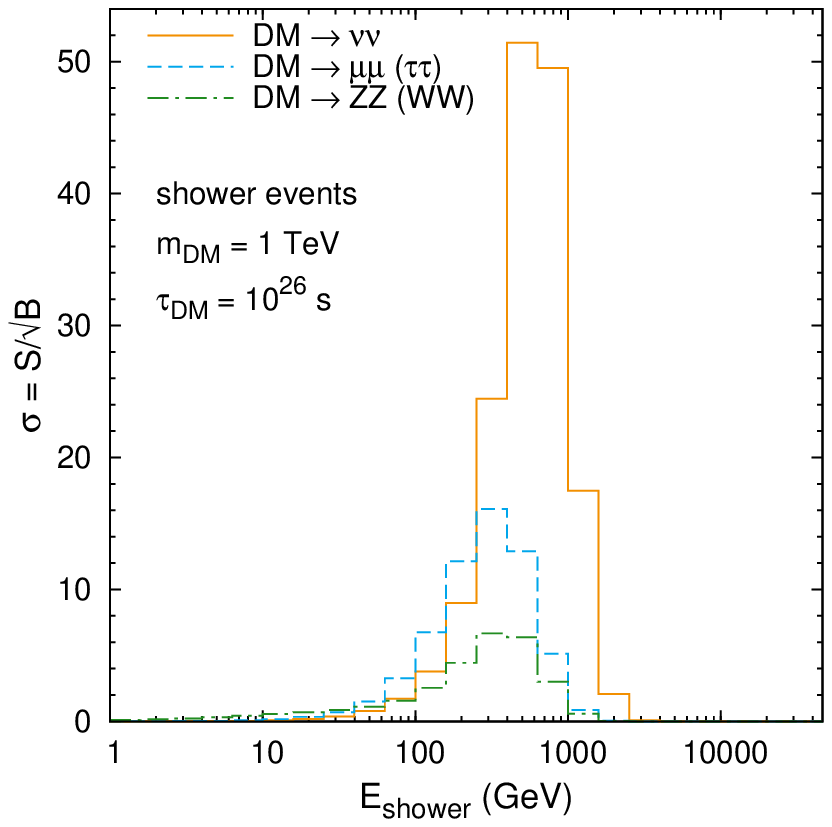}
 \includegraphics[scale=.92,bb=106 57 345 294,clip]{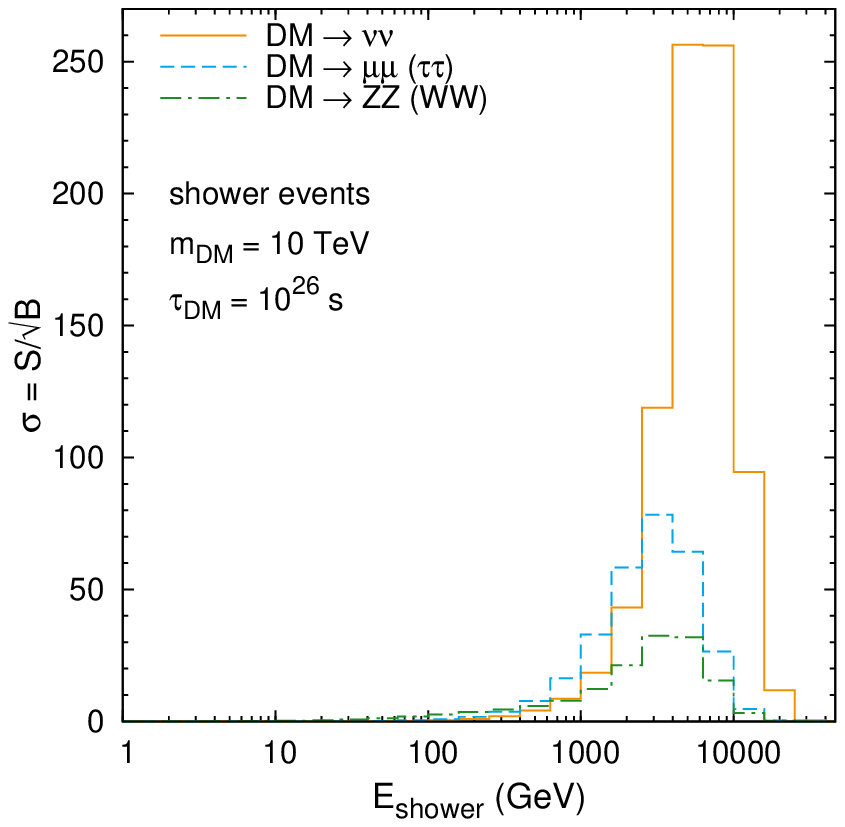}
 \caption{\textit{Top:} Shower rates per km$^3$ of detector volume for the different decay channels 
 of a scalar dark matter candidate compared to the atmospheric background for electromagnetic 
 and hadronic showers. The flux is computed for a dark matter mass of 1\usk TeV (\textit{left}) or 
 10\usk TeV (\textit{right}) and a lifetime of $10^{26}$\usk s using an energy resolution of 0.18 
 in $\log_{10} E$ and five bins per decade. 
 \textit{Bottom:} Statistical significance of the shower signal shown above calculated for every single 
 energy bin using one year of data with an effective volume of 1\usk km$^3$.}
 \label{significance3a}
\end{figure}

\begin{figure}
 \centering
 \includegraphics[scale=.92,bb=103 57 344 294,clip]{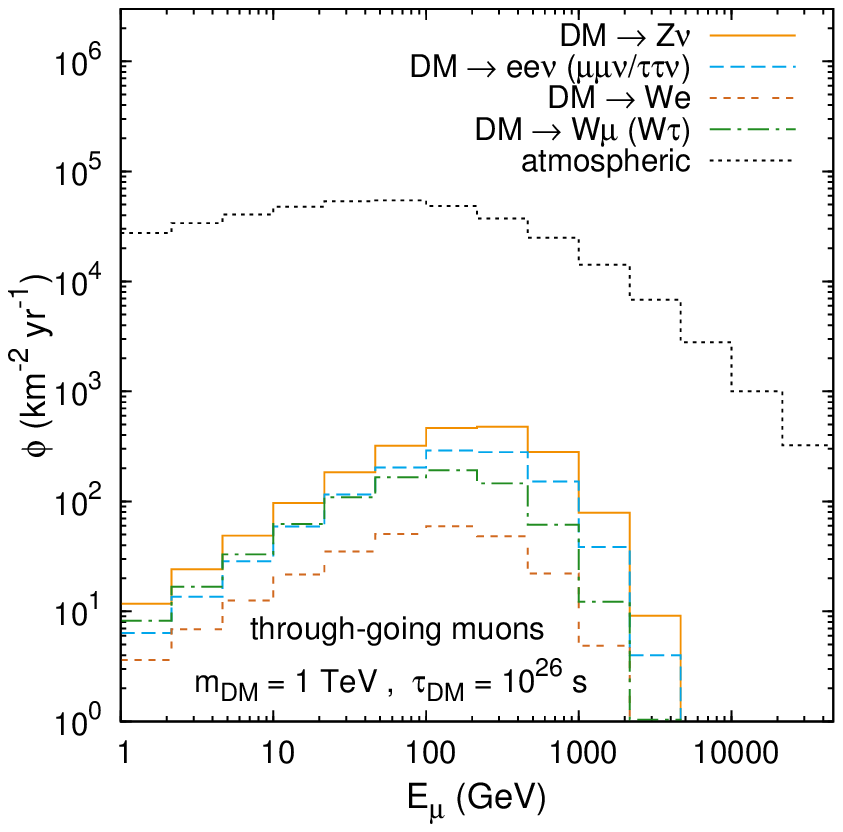}
 \includegraphics[scale=.92,bb=103 57 344 294,clip]{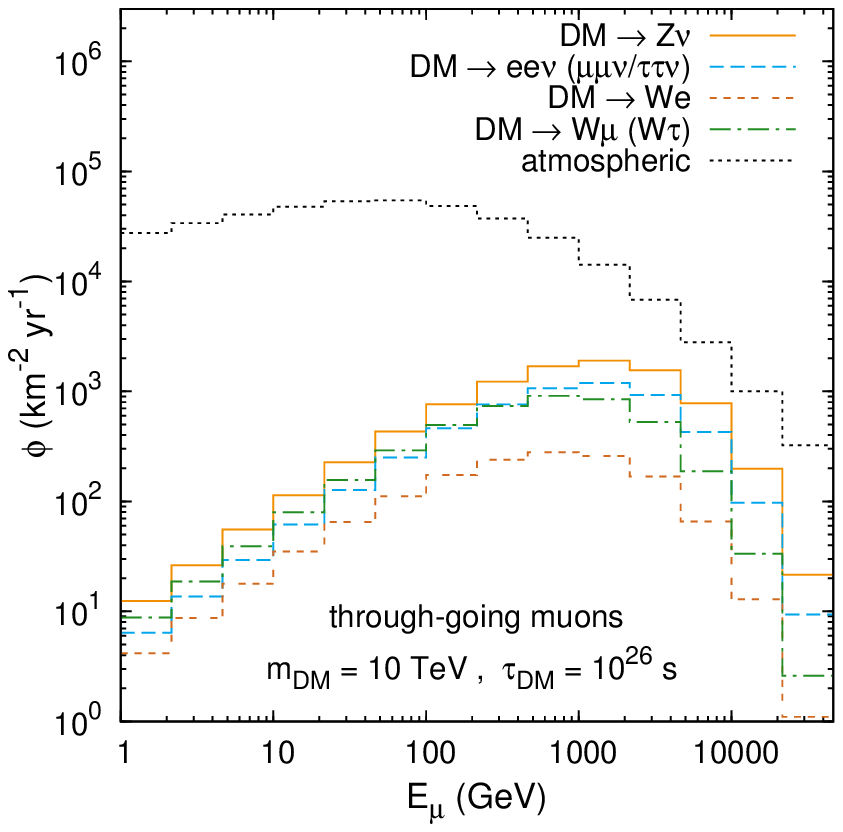}
 \includegraphics[scale=.92,bb=101 57 344 294,clip]{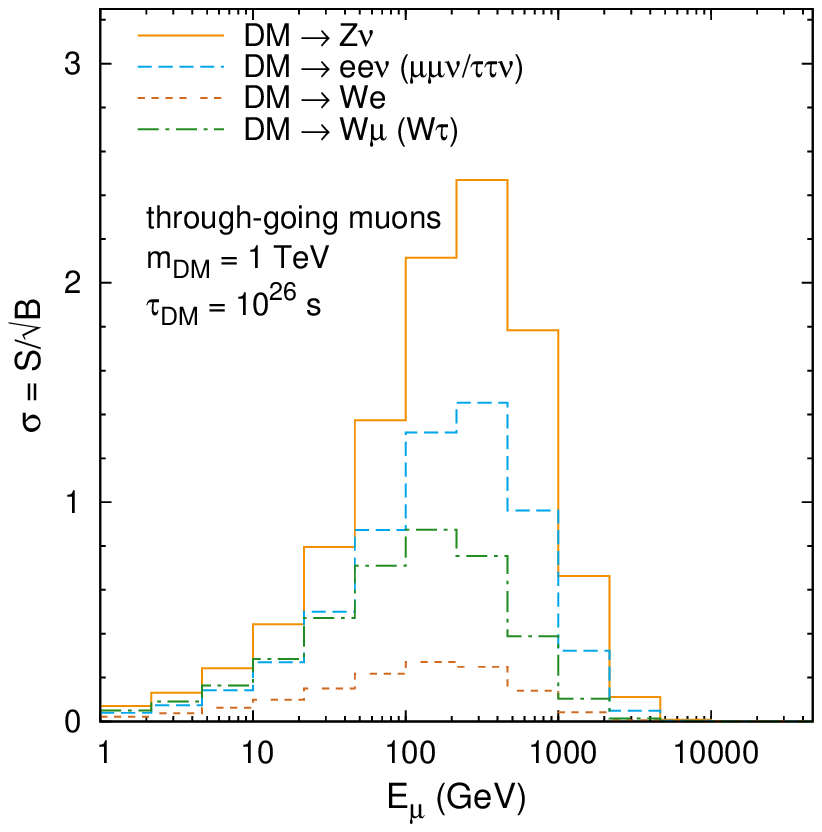}
 \includegraphics[scale=.92,bb=104 57 343 294,clip]{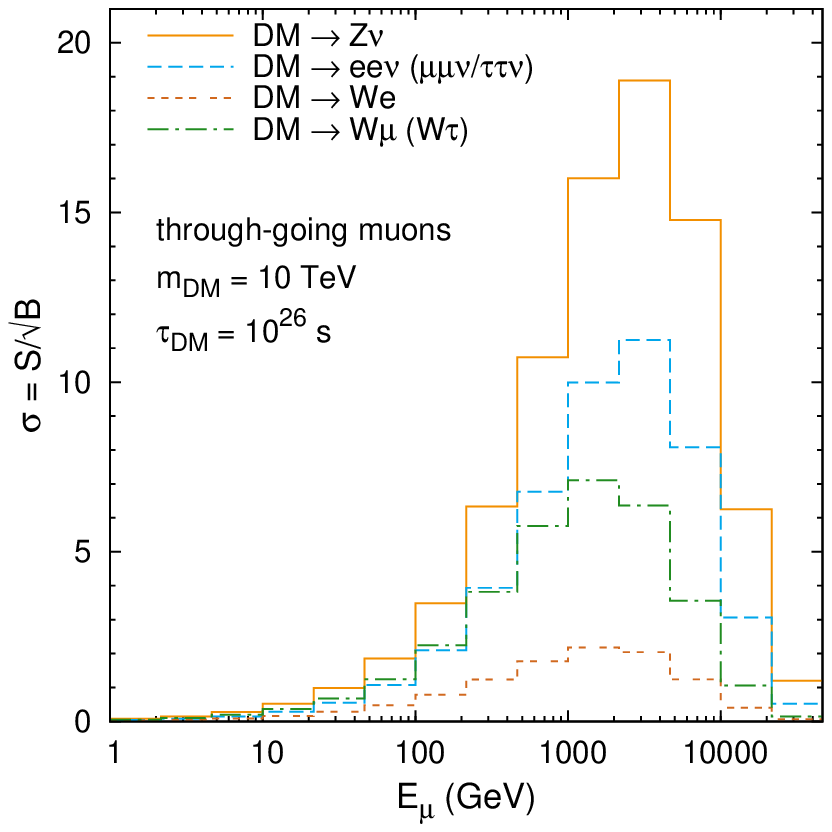}
 \caption{
 \textit{Top:} Muon fluxes for the different decay channels of a fermionic dark matter candidate 
 compared to the atmospheric background for upward through-going muons. 
 The flux is computed for a dark 
 matter mass of 1\usk TeV (\textit{left}) or 10\usk TeV (\textit{right}) and a lifetime of $10^{26}$\usk s 
 using an energy resolution of 0.3 in $\log_{10} E$ and three bins per decade. 
 \textit{Bottom:} Statistical significance of the signal of through-going muons shown above 
 calculated for every single energy bin using one year of data with an effective area of 1\usk km$^2$.}
 \label{significance4a}
\end{figure}

\begin{figure}
 \centering
 \includegraphics[scale=.92,bb=103 57 344 294,clip]{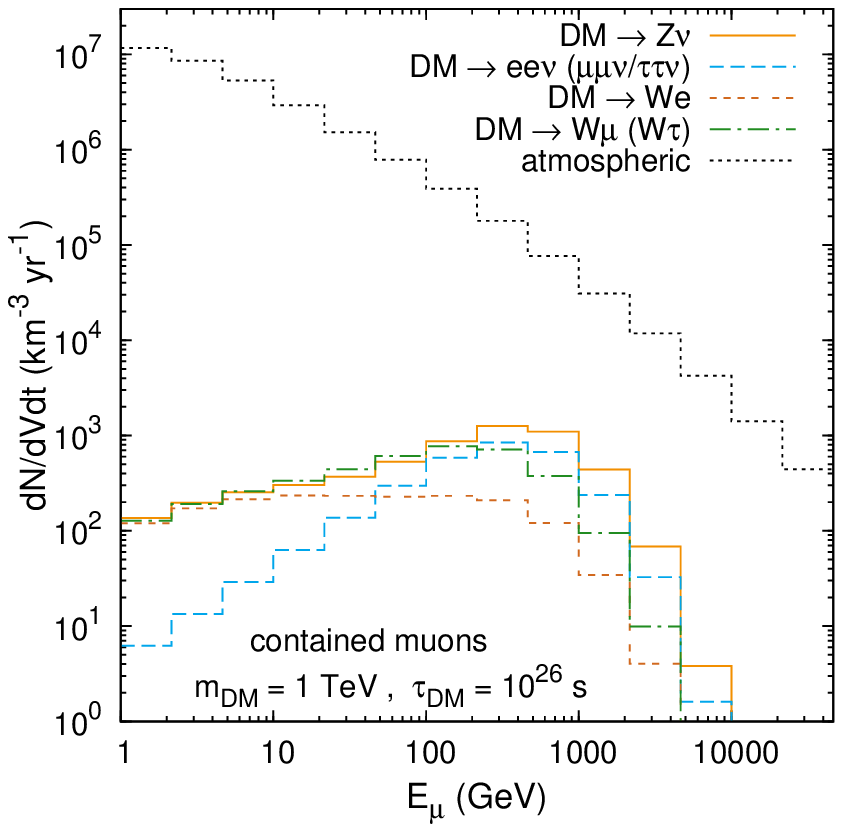}
 \includegraphics[scale=.92,bb=103 57 344 294,clip]{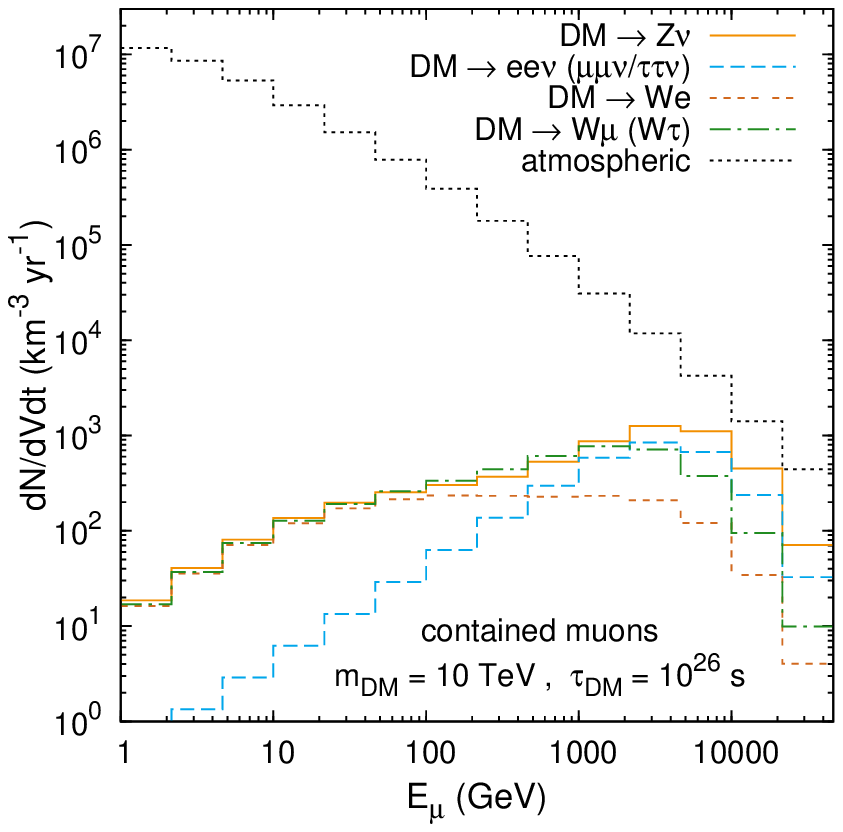}
 \includegraphics[scale=.92,bb=101 57 344 294,clip]{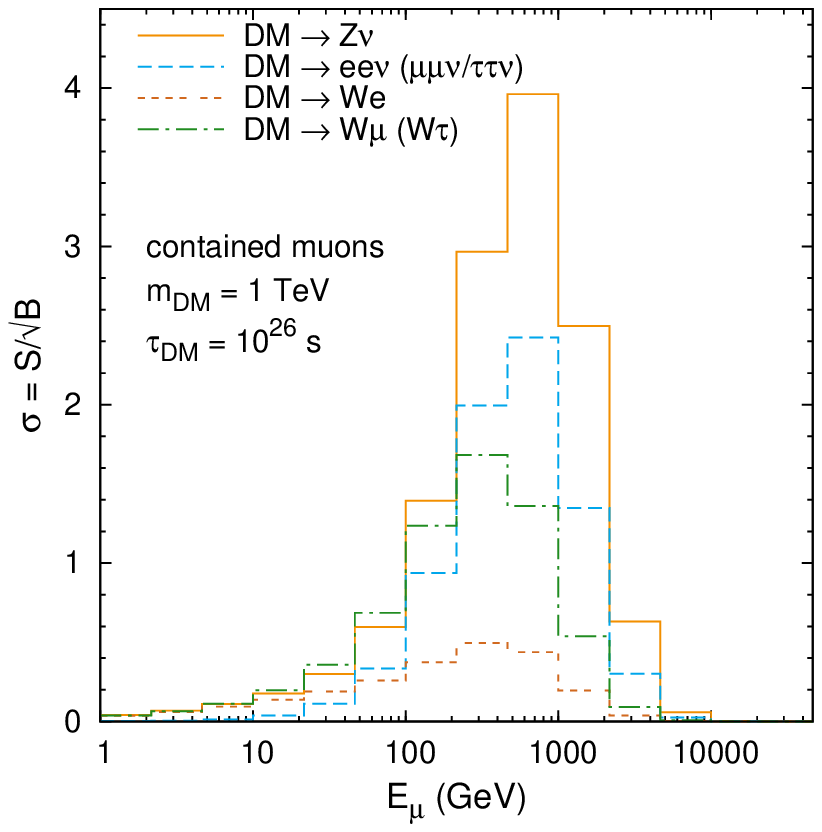}
 \includegraphics[scale=.92,bb=104 57 343 294,clip]{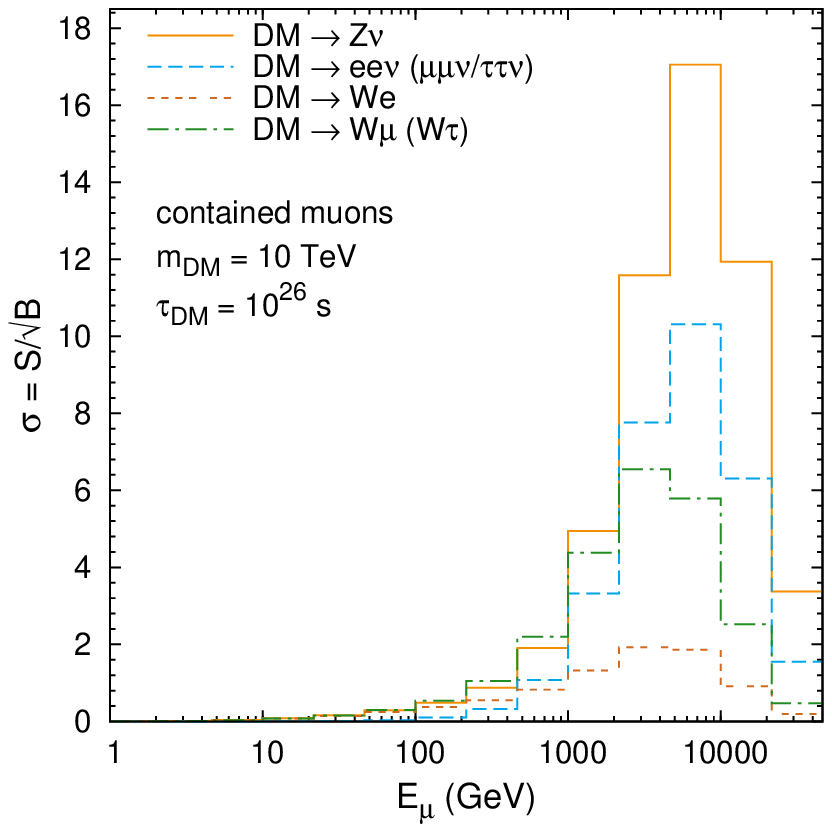}
 \caption{
 \textit{Top:} Muon rates per km$^3$ of detector volume for the different decay channels of a 
 fermionic dark matter candidate compared to the atmospheric background for contained muons. 
 The flux is computed for a dark  matter mass of 1\usk TeV (\textit{left}) or 10\usk TeV (\textit{right}) 
 and a lifetime of 
 $10^{26}$\usk s using an energy resolution of 0.3 in $\log_{10} E$ and three bins per decade. 
 \textit{Bottom:} Statistical significance of the contained muon signal shown above calculated 
 for every single energy bin using one year of data with an effective volume of 1\usk km$^3$.}
 \label{significance5a}
\end{figure}

\begin{figure}
 \centering
 \includegraphics[scale=.92,bb=103 57 344 294,clip]{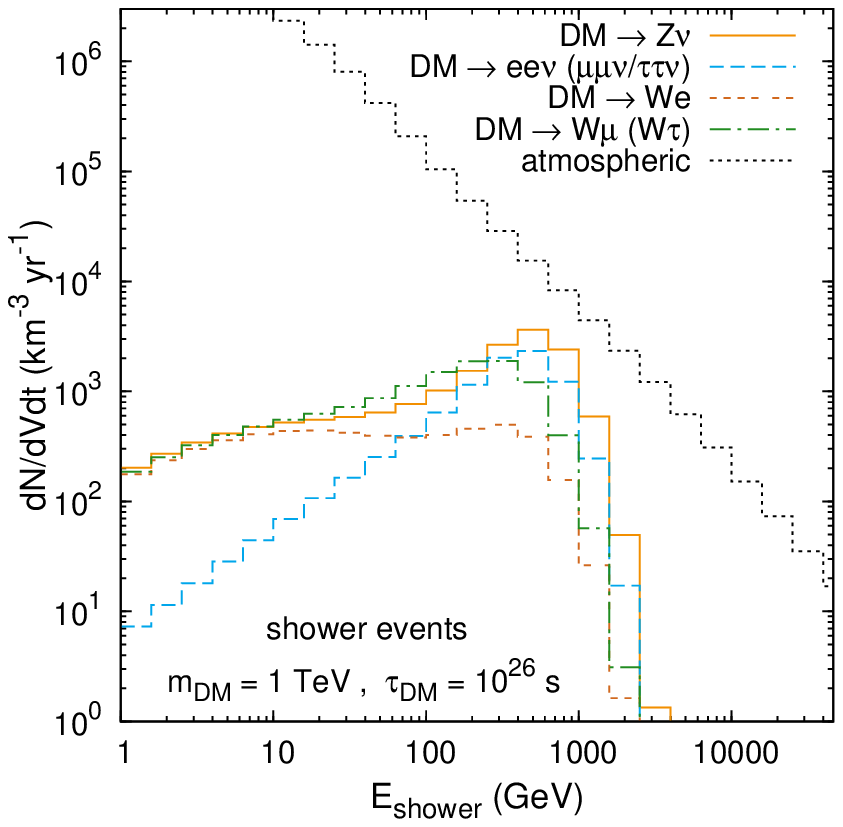}
 \includegraphics[scale=.92,bb=103 57 344 294,clip]{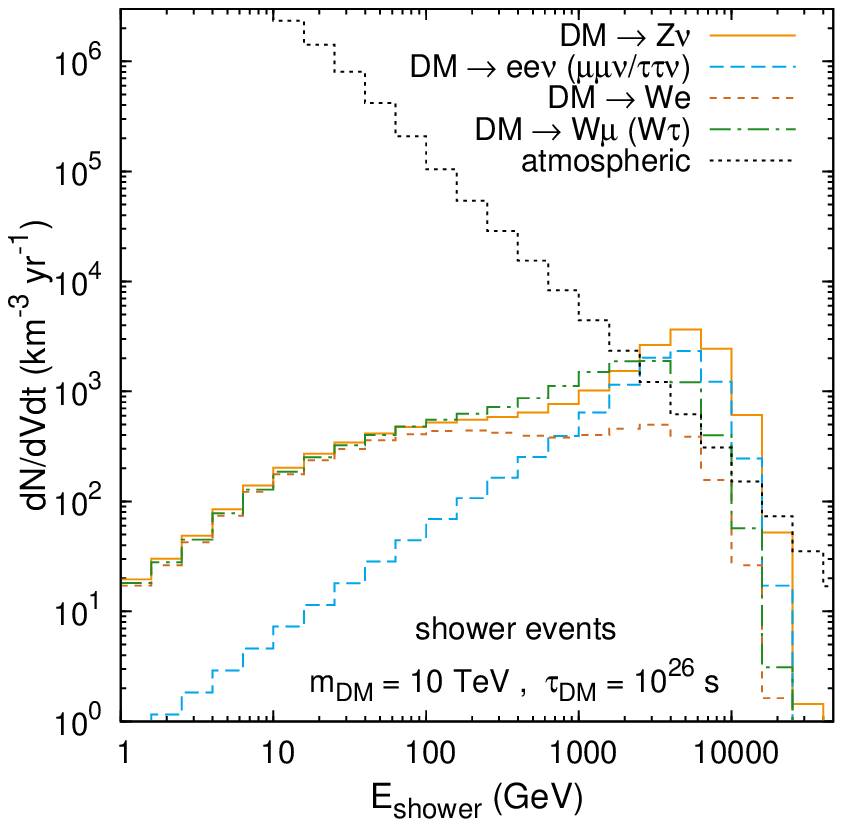}
 \includegraphics[scale=.92,bb=102 57 345 294,clip]{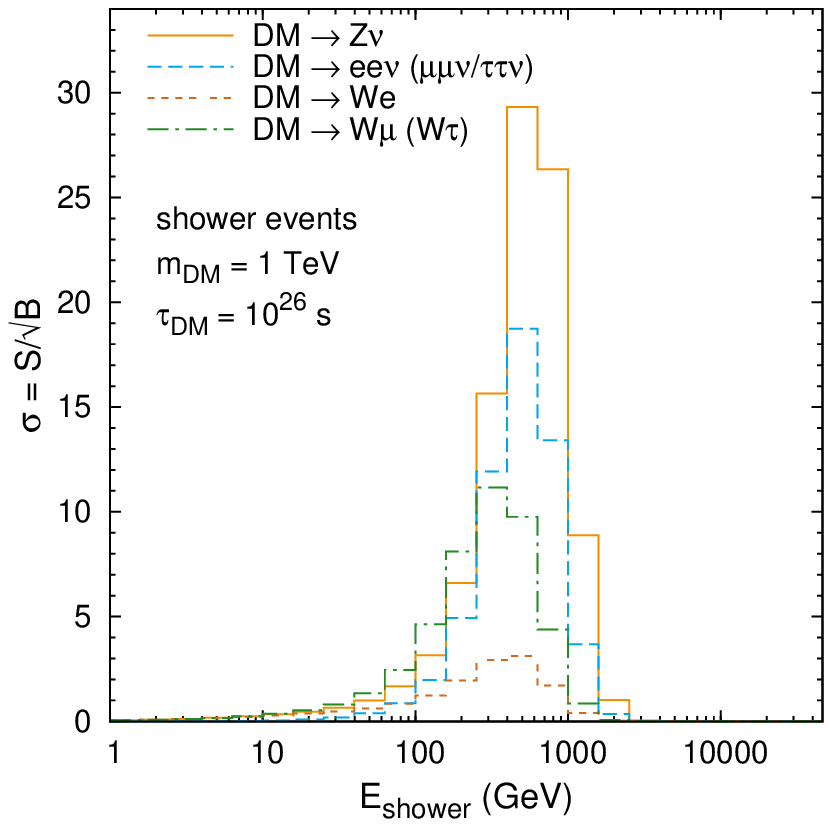}
 \includegraphics[scale=.92,bb=106 57 345 294,clip]{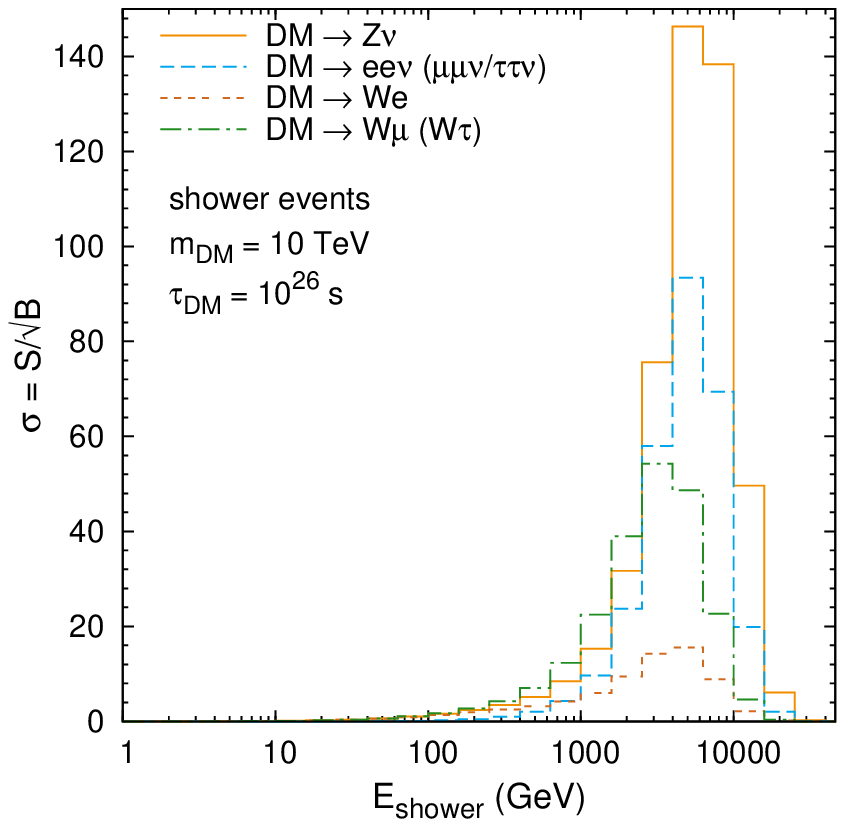}
 \caption{\textit{Top:} Shower rates per km$^3$ of detector volume for the different decay channels 
 of a fermionic dark matter candidate compared to the atmospheric background for electromagnetic 
 and hadronic showers. The flux is computed for a dark matter mass of 1\usk TeV (\textit{left}) or 
 10\usk TeV (\textit{right}) and a lifetime of $10^{26}$\usk s using an energy resolution of 0.18 
 in $\log_{10} E$ and five bins per decade. 
 \textit{Bottom:} Statistical significance of the shower signal shown above calculated for every single 
 energy bin using one year of data with an effective volume of 1\usk km$^3$.}
 \label{significance6a}
\end{figure}

We show in Figures~\ref{significance1a}--\ref{significance6a} the histograms for the signal and the 
atmospheric background using an energy resolution of 0.3 in $\log_{10} E$ and three bins per decade for 
upward through-going and contained muons, and an energy resolution of 0.18 in $\log_{10} E$ and five bins 
per decade for shower events. These figures can be compared to Figures~\ref{SK-Mu-spectra}--\ref{IC-Shower-spectra} 
which show the spectra unbinned and without finite energy resolution. Also shown 
is the significance of the signal over the background in different bins for a 
lifetime of $10^{26}$\usk s for the different channels using one year of data with an effective area of 
1\usk km$^2$ for upward through-going muons and an effective volume of 1\usk km$^3$ for contained muons 
and cascades. All plots are available for both, scalar and fermionic, dark matter candidates.

We see that for nearly any of the spectra, the signal will appear with a large statistical significance 
in more than one single bin and it will be clear that the neutrino signal is not following a power law 
like the atmospheric one. Thus, it is clear that using spectral information it will be possible to set 
much stricter limits on the decaying dark matter parameter space than shown in Figure~\ref{ICprospects}. In 
order to give an idea of the sensitivities that can be obtained using spectral information, we show in 
Table~\ref{sensitivity} the values of the dark matter lifetime for several decay channels that correspond to a 
5 $\sigma$ signal in the most significant energy bin after one year of observation for an idealised detector with 
an effective muon area of 1\usk km$^2$ and an effective volume of 1\usk km$^3$ for contained muons and shower 
events. We see there that the limits from through-going and contained muons are better but not far from those 
shown in Figure~\ref{ICprospects}, while the shower events in principle allow to reach even one order of 
magnitude larger lifetimes. Using not only the dominant energy bin from 
Figures~\ref{significance1a}--\ref{significance6a} but a combination of several energy bins optimised for each 
individual decay channel it will be possible to set even stronger constraints on the dark matter lifetime. So a 
signal in the region preferred by PAMELA should be in the detectable range.

\begin{table}
 \centering
 \begin{tabular}{lccc}
  \hline
  decay channel & through-going muons & contained muons & shower events \\
  \hline
  $DM\rightarrow\nu\bar{\nu}$ & $8.9\times 10^{25}\usk$s & $1.4\times 10^{26}\usk$s & $1.0\times 10^{27}\usk$s \\
  $DM\rightarrow\mu^+\mu^-$ & $2.4\times 10^{25}\usk$s & $4.7\times 10^{25}\usk$s & $3.2\times 10^{26}\usk$s \\
  $DM\rightarrow Z^0Z^0$ & $1.1\times 10^{25}\usk$s & $2.1\times 10^{25}\usk$s & $1.3\times 10^{26}\usk$s \\
  $DM\rightarrow Z^0\nu$ & $4.9\times 10^{25}\usk$s & $7.9\times 10^{25}\usk$s & $5.9\times 10^{26}\usk$s \\
  $DM\rightarrow e^+e^-\nu$ & $2.9\times 10^{25}\usk$s & $4.8\times 10^{25}\usk$s & $3.7\times 10^{26}\usk$s \\
  $DM\rightarrow W^\pm e^\mp$ & $5.4\times 10^{24}\usk$s & $9.9\times 10^{24}\usk$s & $6.2\times 10^{25}\usk$s \\
  $DM\rightarrow W^\pm\mu^\mp$ & $1.7\times 10^{25}\usk$s & $3.4\times 10^{25}\usk$s & $2.2\times 10^{26}\usk$s \\
  \hline
  $DM\rightarrow\nu\bar{\nu}$ & $6.6\times 10^{26}\usk$s & $6.0\times 10^{26}\usk$s & $5.1\times 10^{27}\usk$s \\
  $DM\rightarrow\mu^+\mu^-$ & $1.9\times 10^{26}\usk$s & $1.8\times 10^{26}\usk$s & $1.6\times 10^{27}\usk$s \\
  $DM\rightarrow Z^0Z^0$ & $9.1\times 10^{25}\usk$s & $8.3\times 10^{25}\usk$s & $6.5\times 10^{26}\usk$s \\
  $DM\rightarrow Z^0\nu$ & $3.8\times 10^{26}\usk$s & $3.4\times 10^{26}\usk$s & $2.9\times 10^{27}\usk$s \\
  $DM\rightarrow e^+e^-\nu$ & $2.2\times 10^{26}\usk$s & $2.1\times 10^{26}\usk$s & $1.9\times 10^{27}\usk$s \\
  $DM\rightarrow W^\pm e^\mp$ & $4.4\times 10^{25}\usk$s & $3.9\times 10^{25}\usk$s & $3.1\times 10^{26}\usk$s \\
  $DM\rightarrow W^\pm\mu^\mp$ & $1.4\times 10^{26}\usk$s & $1.3\times 10^{26}\usk$s & $1.1\times 10^{27}\usk$s \\
  \hline
 \end{tabular}
 \caption{Dark matter lifetimes corresponding to a $\sigma=5$ significance in the most significant energy 
bin after one year of observation in an idealised detector with an effective muon area of 1\usk km$^2$ and an 
effective volume for contained muons and showers of 1\usk km$^3$. The numbers are given for dark matter masses of 
1\usk TeV (\textit{top}) and 10\usk TeV (\textit{bottom}). Notice that the sensitivity obtained with 
through-going and contained muons is similar. At larger masses the bound from through-going muons is stronger 
since the statistics increases due to the longer muon range at higher energies. However, neglecting 
reconstruction efficiencies the strongest constraint is obtained from shower events since that channel offers 
the best signal-to-background ratio (see discussion in section~\ref{showers}).}
 \label{sensitivity}
\end{table}

On the other hand, discriminating between the spectra for the different channels will not be so 
straightforward, especially if the mass of the decaying particle is unknown. After convolution with the 
energy resolution, the two-body, three-body or continuum spectra appear quite similar, especially within 
their statistical error, but their significance peaks at slightly different values for the same dark 
matter mass. Note, though, that the signal from a neutrino line remains steeper than the other ones at 
the edge and it may be possible to distinguish it with sufficient statistics. In this respect the more 
promising strategy is probably exploiting the better energy resolution of the shower events, if they can 
be detected. In general a comparison between the different types of events, through-going, contained and 
cascade-like, will make disentangling the shape of the spectra easier. Moreover, if the dark matter mass 
is measured via another channel, like gamma-rays, it may be possible to exploit this information in the 
neutrino fit and compare the position of the neutrino signal ``peak'' in the data with the expectation. 
This should help in disentangling at least a continuum spectrum from the two- and perhaps also three-body 
decay cases. For this specific strategy probably one of the most promising dark matter candidates would be 
a fermion, which may decay into a leptonic three-body final state and subdominantly into $\gamma \nu$, as 
\textit{e.g.} gravitino dark matter with trilinear $R$-parity breaking~\cite{Bomark:2009zm}. Then, if the $\gamma$-channel is 
suppressed by 2--3 orders of magnitude compared to a leptonic three-body decay, the gamma-ray and neutrino 
experiments will actually be exploring the same range of lifetimes. In this case an observation of a gamma 
line by Fermi will provide the dark matter mass measurement and the neutrino signal with a much shorter 
lifetime will point at a three-body or $Z^0\nu$ dominant decay. In case both signals in gamma-rays and 
neutrinos are measured, it may be possible to disentangle also between a scalar and a fermionic dark 
matter candidate, which seems to be very difficult from neutrino measurements alone, since the two types 
of particles produce very similar spectra within the energy resolution of the detectors, as can be seen 
comparing Figures~\ref{significance1a}, \ref{significance2a} and \ref{significance3a} with 
Figures~\ref{significance4a}, \ref{significance5a} and \ref{significance6a}.

\section{Conclusions}
\label{Conclusions}

We have studied in this paper the possible neutrino signals from decaying dark matter, considering 
different decay channels and spectra, both for a scalar and a fermionic candidate. We have concentrated 
here on the region of parameter space that is preferred in order to explain the PAMELA positron excess 
and shown that in this case a signal may soon be visible at neutrino observatories. The non-observation 
of such a signal will put rather strong constraints on the leptophilic decaying dark matter explanation 
of the excess, except for the decay into $e^+e^-$ where no signal in neutrinos is expected. In this sense 
neutrino observations are complementary to other astrophysical constraints coming from radio frequencies 
and inverse Compton emission, which are more sensitive to the electron channel. A neutrino signal will 
allow to disentangle between decaying and annihilating dark matter, by comparing the signal towards and 
away from the Galactic centre~\cite{gamma-direction}, and also between dark matter and astrophysical sources.

More difficult is the identification of the dark matter decay channels, since all neutrino spectra finally 
result in a broad peak in the muon spectrum. However, the analysis of cascade-like events, which contain 
in principle all the neutrino energy and have the advantage of a better energy resolution, may improve 
the situation and allow with sufficient statistics to disentangle at least a line-like feature. Also, the 
neutrino signal alone cannot distinguish between scalar and fermionic dark matter candidates since the 
resulting spectra are very similar in the two cases.

On the other hand, for some of the decay channels discussed here, like the ones with a $Z^0/W^\pm$ gauge 
boson in the final state, corresponding signals are also expected in gamma rays and antiprotons and may 
provide additional information on the model parameters and a cross-check of the decay channel. Even for 
the pure leptophilic channels, gamma rays from final state radiation or from subdominant decays may play 
an important role in discriminating between models due to the better sensitivity in the gamma-ray channel. 
In general, neutrino observations offer complementary information and can be used to test also models 
where the gamma-ray signal is strongly suppressed compared to the leptonic one. The next generation of neutrino 
experiments can therefore be expected to yield some very interesting results.

\section*{Note Added}

During the completion of this work the preprint~\cite{Mandal:2009yk} appeared, presenting a related analysis.

\section*{Acknowledgements}

LC and MG acknowledge the support of the ``Impuls- und Vernetzungsfond'' of the Helmholtz Association under the contract number VH-NG-006 and of the DFG under the Collaborative Research Centre 676. The work of AI and DT was partially supported by the DFG cluster of excellence ``Origin and Structure of the Universe.''


\begin{thebibliography}{99}

\bibitem{DM}
  G.~Bertone, D.~Hooper and J.~Silk,
  %``Particle dark matter: Evidence, candidates and constraints,''
  Phys.\ Rept.\  {\bf 405}, 279 (2005)
  [arXiv:hep-ph/0404175];
  %%CITATION = PRPLC,405,279;%%
  L.~Bergstr\"om,
  %``Non-baryonic dark matter: Observational evidence and detection methods,''
  Rept.\ Prog.\ Phys.\  {\bf 63}, 793 (2000)
  [arXiv:hep-ph/0002126].
  %%CITATION = RPPHA,63,793;%%

\bibitem{lifetime}
  %\cite{PalomaresRuiz:2007ry}
  % \bibitem{PalomaresRuiz:2007ry}
  S.~Palomares-Ruiz,
  %``Model-Independent Bound on the Dark Matter Lifetime,''
  Phys.\ Lett.\  B {\bf 665} (2008) 50
  [arXiv:0712.1937 [astro-ph]];
  %%CITATION = PHLTA,B665,50;%%
  H.~Yuksel and M.~D.~Kistler,
  %``Dark Matter Might Decay... Just Not Today!,''
  Phys.\ Rev.\  D {\bf 78} (2008) 023502
  [arXiv:0711.2906 [astro-ph]];
  %%CITATION = PHRVA,D78,023502;%%
  A.~Boyarsky and O.~Ruchayskiy,
  %``Bounds on Light Dark Matter,''
  arXiv:0811.2385 [astro-ph];
  %%CITATION = ARXIV:0811.2385;%%
  R.~Essig, N.~Sehgal and L.~E.~Strigari,
  %``Bounds on Cross-sections and Lifetimes for Dark Matter Annihilation and
  %Decay into Charged Leptons from Gamma-ray Observations of Dwarf Galaxies,''
  Phys.\ Rev.\  D {\bf 80} (2009) 023506
  [arXiv:0902.4750 [hep-ph]].
  %%CITATION = PHRVA,D80,023506;%%
  
\bibitem{GUT-decay} 
  D.~Eichler,
  %``TeV PARTICLES AS WEAKLY UNSTABLE DARK MATTER,''
  Phys.\ Rev.\ Lett.\  {\bf 63} (1989) 2440;
  %%CITATION = PRLTA,63,2440;%%
  E.~Nardi, F.~Sannino and A.~Strumia,
  %``Decaying Dark Matter can explain the electron/positron excesses,''
  JCAP {\bf 0901} (2009) 043
  [arXiv:0811.4153 [hep-ph]];
  %%CITATION = JCAPA,0901,043;%%
  A.~Arvanitaki, S.~Dimopoulos, S.~Dubovsky, P.~W.~Graham, R.~Harnik and S.~Rajendran,
  %``Astrophysical Probes of Unification,''
  Phys.\ Rev.\  D {\bf 79} (2009) 105022
  [arXiv:0812.2075 [hep-ph]];
  %%CITATION = PHRVA,D79,105022;%%
  K.~Hamaguchi, S.~Shirai and T.~T.~Yanagida,
  %``Cosmic Ray Positron and Electron Excess from Hidden-Fermion Dark Matter
  %Decays,''
  Phys.\ Lett.\  B {\bf 673} (2009) 247
  [arXiv:0812.2374 [hep-ph]].
  %%CITATION = PHLTA,B673,247;%%


\bibitem{R-parity}
  W.~Buchm\"uller, L.~Covi, K.~Hamaguchi, A.~Ibarra and T.~Yanagida,
  %``Gravitino dark matter in R-parity breaking vacua,''
  JHEP {\bf 0703} (2007) 037
  [arXiv:hep-ph/0702184];
  %%CITATION = JHEPA,0703,037;%%
  S.~L.~Chen, R.~N.~Mohapatra, S.~Nussinov and Y.~Zhang,
  %``R-Parity Breaking via Type II Seesaw, Decaying Gravitino Dark Matter and
  %PAMELA Positron Excess,''
  Phys.\ Lett.\  B {\bf 677} (2009) 311
  [arXiv:0903.2562 [hep-ph]];
  %%CITATION = PHLTA,B677,311;%%
  P.~Fileviez Perez and S.~Spinner,
  %``Spontaneous R-Parity Breaking in SUSY Models,''
  Phys.\ Rev.\  D {\bf 80} (2009) 015004
  [arXiv:0904.2213 [hep-ph]];
  %%CITATION = PHRVA,D80,015004;%%
  %\cite{Shirai:2009fq}
  % \bibitem{Shirai:2009fq}
  S.~Shirai, F.~Takahashi and T.~T.~Yanagida,
  %``R-violating Decay of Wino Dark Matter and electron/positron Excesses in the
  %PAMELA/Fermi Experiments,''
  Phys.\ Lett.\  B {\bf 680} (2009) 485
  [arXiv:0905.0388 [hep-ph]];
  %%CITATION = PHLTA,B680,485;%%
  %\cite{Chen:2009mj}
  % \bibitem{Chen:2009mj}
  C.~H.~Chen, C.~Q.~Geng and D.~V.~Zhuridov,
  %``Resolving Fermi, PAMELA and ATIC anomalies in split supersymmetry without
  %R-parity,''
  arXiv:0905.0652 [hep-ph].
  %%CITATION = ARXIV:0905.0652;%%

\bibitem{kin-mix}
  C.~R.~Chen, F.~Takahashi and T.~T.~Yanagida,
  %``Gamma rays and positrons from a decaying hidden gauge boson,''
  Phys.\ Lett.\  B {\bf 671} (2009) 71
  [arXiv:0809.0792 [hep-ph]];
  %%CITATION = PHLTA,B671,71;%%
  A.~Ibarra, A.~Ringwald and C.~Weniger,
  %``Hidden gauginos of an unbroken U(1): Cosmological constraints and
  %phenomenological prospects,''
  JCAP {\bf 0901} (2009) 003
  [arXiv:0809.3196 [hep-ph]].
  %%CITATION = JCAPA,0901,003;%%

\bibitem{DM-decay}
  A.~Ibarra and D.~Tran,
  %``Gamma Ray Spectrum from Gravitino Dark Matter Decay,''
  Phys.\ Rev.\ Lett.\  {\bf 100} (2008) 061301
  [arXiv:0709.4593 [astro-ph]],
  %%CITATION = PRLTA,100,061301;%%
  %``Antimatter Signatures of Gravitino Dark Matter Decay,''
  JCAP {\bf 0807} (2008) 002
  [arXiv:0804.4596 [astro-ph]];
  %%CITATION = JCAPA,0807,002;%%  
  K.~Ishiwata, S.~Matsumoto and T.~Moroi,
  %``High Energy Cosmic Rays from the Decay of Gravitino Dark Matter,''
  Phys.\ Rev.\  D {\bf 78}, 063505 (2008)
  [arXiv:0805.1133 [hep-ph]],
  %%CITATION = PHRVA,D78,063505;%%
  %``High Energy Cosmic Rays from Decaying Supersymmetric Dark Matter,''
  JHEP {\bf 0905}, 110 (2009)
  [arXiv:0903.0242 [hep-ph]];
  %%CITATION = JHEPA,0905,110;%%
  M.~Pospelov and M.~Trott,
  %``R-parity preserving super-WIMP decays,''
  JHEP {\bf 0904}, 044 (2009)
  [arXiv:0812.0432 [hep-ph]];
  %%CITATION = JHEPA,0904,044;%%
  K.~J.~Bae and B.~Kyae,
  %``PAMELA/ATIC Anomaly from Exotic Mediated Dark Matter Decay,''
  JHEP {\bf 0905}, 102 (2009)
  [arXiv:0902.3578 [hep-ph]];
  %%CITATION = JHEPA,0905,102;%%
  K.~Cheung, P.~Y.~Tseng and T.~C.~Yuan,
  %``Double-action dark matter, PAMELA and ATIC,''
  Phys.\ Lett.\  B {\bf 678}, 293 (2009)
  [arXiv:0902.4035 [hep-ph]];
  %%CITATION = PHLTA,B678,293;%%;
  H.~Fukuoka, J.~Kubo and D.~Suematsu,
  %``Anomaly Induced Dark Matter Decay and PAMELA/ATIC Experiments,''
  Phys.\ Lett.\  B {\bf 678}, 401 (2009)
  [arXiv:0905.2847 [hep-ph]];
  %%CITATION = PHLTA,B678,401;%%  
  W.~Buchm\"uller {\it et al.},
  %``Probing Gravitino Dark Matter,''
  JCAP {\bf 0909} (2009) 021
  [arXiv:0906.1187 [hep-ph]];
  %%CITATION = JCAPA,0909,021;%%
  K.~Y.~Choi, D.~E.~Lopez-Fogliani, C.~Munoz and R.~R.~de Austri,
  %``Gamma-ray detection from gravitino dark matter decay in the $\mu\nu$SSM,''
  arXiv:0906.3681 [hep-ph].
  %%CITATION = ARXIV:0906.3681;%%

%\cite{Ibarra:2009dr}
\bibitem{Ibarra:2009dr}
  A.~Ibarra, D.~Tran and C.~Weniger,
  %``Decaying Dark Matter in Light of the PAMELA and Fermi LAT Data,''
  arXiv:0906.1571 [hep-ph], to appear in JCAP.
  %%CITATION = ARXIV:0906.1571;%%

%\cite{Covi:2008jy}
\bibitem{Covi:2008jy}
  L.~Covi, M.~Grefe, A.~Ibarra and D.~Tran,
  %``Unstable Gravitino Dark Matter and Neutrino Flux,''
  JCAP {\bf 0901} (2009) 029
  [arXiv:0809.5030 [hep-ph]];
  %%CITATION = JCAPA,0901,029;%% 
  %\cite{Grefe:2008zz}
  %\bibitem{Grefe:2008zz}
  M.~Grefe,
  %``Neutrino signals from gravitino dark matter with broken R-parity,''
  DESY-THESIS-2008-043.
  %%CITATION = DESY-THESIS-2008-043;%%
  
\bibitem{neutrinos}
  % %\cite{Hisano:2008ah}
  % \bibitem{Hisano:2008ah}
  J.~Hisano, M.~Kawasaki, K.~Kohri and K.~Nakayama,
  %``Neutrino Signals from Annihilating/Decaying Dark Matter in the Light of
  %Recent Measurements of Cosmic Ray Electron/Positron Fluxes,''
  Phys.\ Rev.\  D {\bf 79} (2009) 043516
  [arXiv:0812.0219 [hep-ph]];
  %%CITATION = PHRVA,D79,043516;%%
  % %\cite{Liu:2008ci}
  % \bibitem{Liu:2008ci}
  J.~Liu, P.~f.~Yin and S.~h.~Zhu,
  %``Prospects for Detecting Neutrino Signals from Annihilating/Decaying Dark
  %Matter to Account for the PAMELA and ATIC results,''
  Phys.\ Rev.\  D {\bf 79} (2009) 063522
  [arXiv:0812.0964 [astro-ph]];
  %%CITATION = PHRVA,D79,063522;%%
% %\cite{Hisano:2009fb}
% \bibitem{Hisano:2009fb}
  J.~Hisano, K.~Nakayama and M.~J.~S.~Yang,
  %``Upward muon signals at neutrino detectors as a probe of dark matter
  %properties,''
  Phys.\ Lett.\  B {\bf 678} (2009) 101
  [arXiv:0905.2075 [hep-ph]];
  %%CITATION = PHLTA,B678,101;%%
% %\cite{Buckley:2009kw}
% \bibitem{Buckley:2009kw}
  M.~R.~Buckley, K.~Freese, D.~Hooper, D.~Spolyar and H.~Murayama,
  %``High-Energy Neutrino Signatures of Dark Matter Decaying into Leptons,''
  arXiv:0907.2385 [astro-ph.HE];
  %%CITATION = ARXIV:0907.2385;%%
  %\cite{Liu:2009ac}
% \bibitem{Liu:2009ac}
  J.~Liu, Q.~Yuan, X.~Bi, H.~Li and X.~Zhang,
  %``Neutrino emission from dark matter annihilation/decay in light of cosmic
  %$e^{\pm}$ and $\bar{p}$ data,''
  arXiv:0911.1002 [astro-ph.CO].
  %%CITATION = ARXIV:0911.1002;%%

\bibitem{leptophilic}
  P.~f.~Yin, Q.~Yuan, J.~Liu, J.~Zhang, X.~j.~Bi and S.~h.~Zhu,
  %``PAMELA data and leptonically decaying dark matter,''
  Phys.\ Rev.\  D {\bf 79}, 023512 (2009)
  [arXiv:0811.0176 [hep-ph]];
  %%CITATION = PHRVA,D79,023512;%%
  %\cite{Kyae:2009jt}
% \bibitem{Kyae:2009jt}
  B.~Kyae,
  %``PAMELA/ATIC anomaly from the meta-stable extra dark matter component and
  %the leptophilic Yukawa interaction,''
  JCAP {\bf 0907} (2009) 028
  [arXiv:0902.0071 [hep-ph]];
  %%CITATION = JCAPA,0907,028;%%
  A.~Ibarra, A.~Ringwald, D.~Tran and C.~Weniger,
  %``Cosmic Rays from Leptophilic Dark Matter Decay via Kinetic Mixing,''
  JCAP {\bf 0908} (2009) 017
  [arXiv:0903.3625 [hep-ph]];
  %%CITATION = JCAPA,0908,017;%%
  %\cite{Fukuoka:2009cu}
% \bibitem{Fukuoka:2009cu}
  H.~Fukuoka, J.~Kubo and D.~Suematsu,
  %``Anomaly Induced Dark Matter Decay and PAMELA/ATIC Experiments,''
  Phys.\ Lett.\  B {\bf 678} (2009) 401
  [arXiv:0905.2847 [hep-ph]];
  %%CITATION = PHLTA,B678,401;%%
  %\cite{Chen:2009zpa}
% \bibitem{Chen:2009zpa}
  C.~H.~Chen,
  %``Resolution to neutrino masses, baryon asymmetry in leptogenesis and
  %cosmic-ray anomalies,''
  arXiv:0905.3425 [hep-ph];
  %%CITATION = ARXIV:0905.3425;%%
  %\cite{Demir:2009kc}
% \bibitem{Demir:2009kc}
  D.~A.~Demir, L.~L.~Everett, M.~Frank, L.~Selbuz and I.~Turan,
  %``Sneutrino Dark Matter: Symmetry Protection and Cosmic Ray Anomalies,''
  arXiv:0906.3540 [hep-ph];
  %%CITATION = ARXIV:0906.3540;%%
  %\cite{Ruderman:2009tj}
% \bibitem{Ruderman:2009tj}
  J.~T.~Ruderman and T.~Volansky,
  %``Decaying into the Hidden Sector,''
  arXiv:0908.1570 [hep-ph].
  %%CITATION = ARXIV:0908.1570;%%

\bibitem{PAMELA-decay}
% %\cite{Ibarra:2008jk}
% \bibitem{Ibarra:2008jk}
  A.~Ibarra and D.~Tran,
  %``Decaying Dark Matter and the PAMELA Anomaly,''
  JCAP {\bf 0902} (2009) 021
  [arXiv:0811.1555 [hep-ph]].
  %%CITATION = JCAPA,0902,021;%%

%\cite{Adriani:2008zr}
\bibitem{Adriani:2008zr}
  O.~Adriani {\it et al.}  [PAMELA Collaboration],
  %``An anomalous positron abundance in cosmic rays with energies 1.5.100 GeV,''
  Nature {\bf 458} (2009) 607
  [arXiv:0810.4995 [astro-ph]].
  %%CITATION = NATUA,458,607;%%

\bibitem{FermiHessAtic}
% %\cite{Abdo:2009zk}
% \bibitem{Abdo:2009zk}
  A.~A.~Abdo {\it et al.}  [The Fermi LAT Collaboration],
  %``Measurement of the Cosmic Ray e+ plus e- spectrum from 20 GeV to 1 TeV with
  %the Fermi Large Area Telescope,''
  Phys.\ Rev.\ Lett.\  {\bf 102} (2009) 181101
  [arXiv:0905.0025 [astro-ph.HE]];
  %%CITATION = PRLTA,102,181101;%%
% %\cite{Collaboration:2008aaa}
% \bibitem{Collaboration:2008aaa}
  F.~Aharonian {\it et al.}  [H.E.S.S. Collaboration],
  %``The energy spectrum of cosmic-ray electrons at TeV energies,''
  Phys.\ Rev.\ Lett.\  {\bf 101} (2008) 261104
  [arXiv:0811.3894 [astro-ph]],
  %%CITATION = PRLTA,101,261104;%%
% %\cite{Aharonian:2009ah}
% \bibitem{Aharonian:2009ah}
%   F.~Aharonian {\it et al.}  [H.E.S.S. Collaboration],
  %``Probing the ATIC peak in the cosmic-ray electron spectrum with H.E.S.S,''
  arXiv:0905.0105 [astro-ph.HE];
  %%CITATION = ARXIV:0905.0105;%%
% %\cite{:2008zzr}
% \bibitem{:2008zzr}
  J.~Chang {\it et al.},
  %``An Excess Of Cosmic Ray Electrons At Energies Of 300.800 Gev,''
  Nature {\bf 456} (2008) 362.
  %%CITATION = NATUA,456,362;%%

%\cite{Adriani:2008zq}
\bibitem{Adriani:2008zq}
  O.~Adriani {\it et al.},
  %``A new measurement of the antiproton-to-proton flux ratio up to 100 GeV in
  %the cosmic radiation,''
  Phys.\ Rev.\ Lett.\  {\bf 102} (2009) 051101
  [arXiv:0810.4994 [astro-ph]].
  %%CITATION = PRLTA,102,051101;%%
  
%\cite{Hooper:2009cs}
 \bibitem{Hooper:2009cs}
 D.~Hooper and K.~M.~Zurek,
 %``Pamela, FGST and Sub-Tev Dark Matter,''
 arXiv:0909.4163 [hep-ph].
 %%CITATION = ARXIV:0909.4163;%%

\bibitem{Radio}
  G.~Bertone, M.~Cirelli, A.~Strumia and M.~Taoso,
  %``Gamma-ray and radio tests of the e+e- excess from DM annihilations,''
  JCAP {\bf 0903}, 009 (2009)
  [arXiv:0811.3744 [astro-ph]];
  %%CITATION = JCAPA,0903,009;%%
  L.~Bergstr\"om, G.~Bertone, T.~Bringmann, J.~Edsj\"o and M.~Taoso,
  %``Gamma-ray and Radio Constraints of High Positron Rate Dark Matter Models
  %Annihilating into New Light Particles,''
  Phys.\ Rev.\  D {\bf 79} (2009) 081303
  [arXiv:0812.3895 [astro-ph]];
  %\cite{Ishiwata:2008qy}
% \bibitem{Ishiwata:2008qy}
  K.~Ishiwata, S.~Matsumoto and T.~Moroi,
  %``Synchrotron Radiation from the Galactic Center in Decaying Dark Matter
  %Scenario,''
  Phys.\ Rev.\  D {\bf 79} (2009) 043527
  [arXiv:0811.4492 [astro-ph]];
  %%CITATION = PHRVA,D79,043527;%%
  %%CITATION = PHRVA,D79,081303;%%
  L.~Zhang, G.~Sigl and J.~Redondo,
  %``Galactic Signatures of Decaying Dark Matter,''
  JCAP {\bf 0909} (2009) 012
  [arXiv:0905.4952 [astro-ph.GA]].
  %%CITATION = JCAPA,0909,012;%%
   
\bibitem{IC}
  J.~Zhang, X.~J.~Bi, J.~Liu, S.~M.~Liu, P.~F.~Yin, Q.~Yuan and S.~H.~Zhu,
  %``Discriminating different scenarios to account for the cosmic e$^\pm$ excess
  %by synchrotron and inverse Compton radiation,''
  Phys.\ Rev.\  D {\bf 80} (2009) 023007
  [arXiv:0812.0522 [astro-ph]];
  %%CITATION = PHRVA,D80,023007;%%
  M.~Cirelli and P.~Panci,
  %``Inverse Compton constraints on the Dark Matter e+e- excesses,''
  Nucl.\ Phys.\  B {\bf 821} (2009) 399
  [arXiv:0904.3830 [astro-ph.CO]];
  %%CITATION = NUPHA,B821,399;%%
  S.~Profumo and T.~E.~Jeltema,
  %``Extragalactic Inverse Compton Light from Dark Matter Annihilation and the
  %Pamela Positron Excess,''
  JCAP {\bf 0907} (2009) 020
  [arXiv:0906.0001 [astro-ph.CO]];
  %%CITATION = JCAPA,0907,020;%%
  %\cite{Ishiwata:2009dk}
  % \bibitem{Ishiwata:2009dk}
  K.~Ishiwata, S.~Matsumoto and T.~Moroi,
  %``Cosmic Gamma-ray from Inverse Compton Process in Unstable Dark Matter
  %Scenario,''
  Phys.\ Lett.\  B {\bf 679} (2009) 1
  [arXiv:0905.4593 [astro-ph.CO]].
  %%CITATION = PHLTA,B679,1;%%
  
\bibitem{astro-positron}
  S.~Profumo,
  %``Dissecting Pamela (and ATIC) with Occam's Razor: existing, well-known
  %Pulsars naturally account for the 'anomalous' Cosmic-Ray Electron and
  %Positron Data,''
  arXiv:0812.4457 [astro-ph];
  %%CITATION = ARXIV:0812.4457;%%
  N.~J.~Shaviv, E.~Nakar and T.~Piran,
  %``Natural explanation for the anomalous positron to electron ratio with
  %supernova remnants as the sole cosmic ray source,''
  Phys.\ Rev.\ Lett.\  {\bf 103} (2009) 111302
  [arXiv:0902.0376 [astro-ph.HE]];
  %%CITATION = PRLTA,103,111302;%%
  P.~Blasi and P.~D.~Serpico,
  %``High-energy antiprotons from old supernova remnants,''
  Phys.\ Rev.\ Lett.\  {\bf 103} (2009) 081103
  [arXiv:0904.0871 [astro-ph.HE]];
  %%CITATION = PRLTA,103,081103;%%
  V.~Barger, Y.~Gao, W.~Y.~Keung, D.~Marfatia and G.~Shaughnessy,
  %``Dark matter and pulsar signals for Fermi LAT, PAMELA, ATIC, HESS and WMAP
  %data,''
  Phys.\ Lett.\  B {\bf 678} (2009) 283
  [arXiv:0904.2001 [hep-ph]];
  %%CITATION = PHLTA,B678,283;%%
  D.~Grasso {\it et al.}  [FERMI-LAT Collaboration],
  %``On possible interpretations of the high energy electron-positron spectrum
  %measured by the Fermi Large Area Telescope,''
  Astropart.\ Phys.\  {\bf 32} (2009) 140
  [arXiv:0905.0636 [astro-ph.HE]].
  %%CITATION = APHYE,32,140;%%
  
\bibitem{gamma-direction}
%\cite{Bertone:2007aw}
%\bibitem{Bertone:2007aw}
  G.~Bertone, W.~Buchm\"uller, L.~Covi and A.~Ibarra,
  %``Gamma-Rays from Decaying Dark Matter,''
  JCAP {\bf 0711} (2007) 003
  [arXiv:0709.2299 [astro-ph]];
  %%CITATION = JCAPA,0711,003;%%
% %\cite{Ibarra:2009nw}
% \bibitem{Ibarra:2009nw}
  A.~Ibarra, D.~Tran and C.~Weniger,
  %``Detecting Gamma-Ray Anisotropies from Decaying Dark Matter: Prospects for
  %Fermi LAT,''
  arXiv:0909.3514 [hep-ph], to appear in Phys.\ Rev.\  D.
  %%CITATION = ARXIV:0909.3514;%%  
  
%\cite{Strumia:2006db}
\bibitem{Strumia:2006db}
  A.~Strumia and F.~Vissani,
  %``Neutrino masses and mixings and.,''
  arXiv:hep-ph/0606054.
  %%CITATION = HEP-PH/0606054;%%
  
%\cite{Schwetz:2008er}
\bibitem{Schwetz:2008er}
  T.~Schwetz, M.~Tortola and J.~W.~F.~Valle,
  %``Three-flavour neutrino oscillation update,''
  New J.\ Phys.\  {\bf 10} (2008) 113011
  [arXiv:0808.2016 [hep-ph]].
  %%CITATION = NJOPF,10,113011;%%
  
%\cite{Honda:2006qj}
\bibitem{Honda:2006qj}
  M.~Honda, T.~Kajita, K.~Kasahara, S.~Midorikawa and T.~Sanuki,
  %``Calculation of atmospheric neutrino flux using the interaction model
  %calibrated with atmospheric muon data,''
  Phys.\ Rev.\  D {\bf 75} (2007) 043006
  [arXiv:astro-ph/0611418].
  %%CITATION = PHRVA,D75,043006;%%
  
%\cite{Volkova:2001th}
\bibitem{Volkova:2001th}
  L.~V.~Volkova and G.~T.~Zatsepin,
  %``Fluxes of cosmic ray muons and atmospheric neutrinos at high energies,''
  Phys.\ Atom.\ Nucl.\  {\bf 64} (2001) 266
  [Yad.\ Fiz.\  {\bf 64} (2001) 313].
  %%CITATION = YAFIA,64,313;%%
  
%\cite{Pasquali:1998xf}
\bibitem{Pasquali:1998xf}
  L.~Pasquali and M.~H.~Reno,
  %``Tau neutrino fluxes from atmospheric charm,''
  Phys.\ Rev.\  D {\bf 59} (1999) 093003
  [arXiv:hep-ph/9811268].
  %%CITATION = PHRVA,D59,093003;%%
  
%\cite{Ingelman:1996mj}
\bibitem{Ingelman:1996mj}
  G.~Ingelman and M.~Thunman,
  %``High Energy Neutrino Production by Cosmic Ray Interactions in the Sun,''
  Phys.\ Rev.\  D {\bf 54}, 4385 (1996)
  [arXiv:hep-ph/9604288].
  %%CITATION = PHRVA,D54,4385;%%
  
%\cite{Athar:2004um}
\bibitem{Athar:2004um}
  H.~Athar, F.~F.~Lee and G.~L.~Lin,
  %``Tau neutrino astronomy in GeV energies,''
  Phys.\ Rev.\  D {\bf 71}, 103008 (2005)
  [arXiv:hep-ph/0407183];
  %%CITATION = PHRVA,D71,103008;%%
% %\cite{Ingelman:1996md}
% \bibitem{Ingelman:1996md}
  G.~Ingelman and M.~Thunman,
  %``Particle Production in the Interstellar Medium,''
  arXiv:hep-ph/9604286.
  %%CITATION = HEP-PH/9604286;%%
  
%\cite{Sjostrand:2006za}
\bibitem{Sjostrand:2006za}
  T.~Sj\"ostrand, S.~Mrenna and P.~Skands,
  %``PYTHIA 6.4 Physics and Manual,''
  JHEP {\bf 0605} (2006) 026
  [arXiv:hep-ph/0603175].
  %%CITATION = JHEPA,0605,026;%%
  
%\cite{Daum:1994bf}
\bibitem{Daum:1994bf}
  K.~Daum {\it et al.}  [Fr\'ejus Collaboration.],
  %``Determination of the atmospheric neutrino spectra with the Frejus
  %detector,''
  Z.\ Phys.\  C {\bf 66} (1995) 417.
  %%CITATION = ZEPYA,C66,417;%%
  
%\cite{GonzalezGarcia:2006ay}
\bibitem{GonzalezGarcia:2006ay}
  M.~C.~Gonzalez-Garcia, M.~Maltoni and J.~Rojo,
  %``Determination of the atmospheric neutrino fluxes from atmospheric  neutrino
  %data,''
  JHEP {\bf 0610} (2006) 075
  [arXiv:hep-ph/0607324].
  %%CITATION = JHEPA,0610,075;%%

%\cite{Collaboration:2009nf}
\bibitem{Collaboration:2009nf}
  R.~Abbasi {\it et al.}  [IceCube Collaboration], 
  %``Determination of the Atmospheric Neutrino Flux and Searches for New Physics
  %with AMANDA-II,''
  Phys.\ Rev.\  D {\bf 79} (2009) 102005
  [arXiv:0902.0675 [astro-ph.HE]]; 
  %%CITATION = PHRVA,D79,102005;%% 
%\cite{:2007td}
%\bibitem{:2007td}
  K.~M\"unich and J.~L\"unemann for~the~IceCube~Collaboration,
  %``The IceCube Collaboration: contributions to the 30th International Cosmic
  %Ray Conference (ICRC 2007),''
  arXiv:0711.0353 [astro-ph].
  %%CITATION = ARXIV:0711.0353;%%
  
%\cite{Chirkin:2009}
\bibitem{Chirkin:2009}
  D.~Chirkin for~the~IceCube~Collaboration, \\
  available at http://www.srl.utu.fi/AuxDOC/kocharov/ICRC2009/pdf/icrc1418.pdf
  
%\cite{Barger:2007xf}
\bibitem{Barger:2007xf}
  V.~Barger, W.~Y.~Keung, G.~Shaughnessy and A.~Tregre,
  %``High energy neutrinos from neutralino annihilations in the Sun,''
  Phys.\ Rev.\  D {\bf 76} (2007) 095008
  [arXiv:0708.1325 [hep-ph]].
  %%CITATION = PHRVA,D76,095008;%%

%\cite{Lohmann:1985qg}
\bibitem{Lohmann:1985qg}
  W.~Lohmann, R.~Kopp and R.~Voss,
  %``Energy Loss Of Muons In The Energy Range 1-Gev To 10000-Gev,''
  CERN Report 85-03 (1985);
  %%CITATION = CERN-YELLOW-85-03;%%
%\cite{Groom:2001kq}
%\bibitem{Groom:2001kq}
  D.~E.~Groom, N.~V.~Mokhov and S.~I.~Striganov,
  %``Muon stopping power and range tables 10-MeV to 100-TeV,''
  Atom.\ Data Nucl.\ Data Tabl.\  {\bf 78} (2001) 183.\\
  %%CITATION = ADNDA,78,183;%%
  Tables for muon energy loss are available at \\http://pdg.lbl.gov/2009/AtomicNuclearProperties/.
  
%\cite{Erkoca:2009by}
\bibitem{Erkoca:2009by}
  A.~E.~Erkoca, M.~H.~Reno and I.~Sarcevic,
  %``Muon Fluxes From Dark Matter Annihilation,''
  Phys.\ Rev.\  D {\bf 80} (2009) 043514
  [arXiv:0906.4364 [hep-ph]].
  %%CITATION = PHRVA,80,043514;%%
  
%\cite{Schonert:2008is}
\bibitem{Schonert:2008is}
  S.~Schonert, T.~K.~Gaisser, E.~Resconi and O.~Schulz,
  %``Vetoing atmospheric neutrinos in a high energy neutrino telescope,''
  Phys.\ Rev.\  D {\bf 79} (2009) 043009
  [arXiv:0812.4308 [astro-ph]].
  %%CITATION = PHRVA,D79,043009;%%
  
%\cite{Cowen:2007ny}
\bibitem{Cowen:2007ny}
  D.~F.~Cowen  [IceCube Collaboration],
  %``Tau neutrinos in IceCube,''
  J.\ Phys.\ Conf.\ Ser.\  {\bf 60} (2007) 227.
  %%CITATION = 00462,60,227;%%
  
%\cite{D'Agostino:2009sj}
\bibitem{D'Agostino:2009sj}
  M.~D'Agostino for~the~IceCube~Collaboration,
  %``A Search For Atmospheric Neutrino-Induced Cascades with IceCube,''
  arXiv:0910.0215 [astro-ph.HE].
  %%CITATION = ARXIV:0910.0215;%%
  
%\cite{Desai:2004pq}
\bibitem{Desai:2004pq}
  S.~Desai {\it et al.}  [Super-Kamiokande Collaboration],
  %``Search for dark matter WIMPs using upward through-going muons in
  %Super-Kamiokande,''
  Phys.\ Rev.\  D {\bf 70} (2004) 083523
  [Erratum-ibid.\  D {\bf 70} (2004) 109901]
  [arXiv:hep-ex/0404025].
  %%CITATION = PHRVA,D70,083523;%%
  
%\cite{Montaruli:2009kv}
\bibitem{Montaruli:2009kv}
  T.~Montaruli,
  %``Review on Neutrino Telescopes,''
  Nucl.\ Phys.\ Proc.\ Suppl.\  {\bf 190} (2009) 101
  [arXiv:0901.2661 [astro-ph]].
  %%CITATION = NUPHZ,190,101;%%

%\cite{Wiebusch:2009jf}
\bibitem{Wiebusch:2009jf}
  C.~Wiebusch for~the~IceCube~Collaboration,
  %``Physics Capabilities of the IceCube DeepCore Detector,''
  arXiv:0907.2263 [astro-ph.IM].
  %%CITATION = ARXIV:0907.2263;%%
  
%\cite{Resconi:2008fe}
\bibitem{Resconi:2008fe}
  E.~Resconi for~the~IceCube~Collaboration,
  %``Status and prospects of the IceCube neutrino telescope,''
  Nucl.\ Instrum.\ Meth.\  A {\bf 602} (2009) 7
  [arXiv:0807.3891 [astro-ph]].
  %%CITATION = NUIMA,A602,7;%%
  
%\cite{Bomark:2009zm}
\bibitem{Bomark:2009zm}
  N.~E.~Bomark, S.~Lola, P.~Osland and A.~R.~Raklev,
  %``Photon, Neutrino and Charged Particle Spectra from R-violating Gravitino
  %Decays,''
  arXiv:0911.3376 [hep-ph].
  %%CITATION = ARXIV:0911.3376;%%
  
%\cite{Mandal:2009yk}
\bibitem{Mandal:2009yk}
  S.~K.~Mandal, M.~R.~Buckley, K.~Freese, D.~Spolyar and H.~Murayama,
  %``Cascade Events at IceCube+DeepCore as a Definitive Constraint on the Dark
  %Matter Interpretation of the PAMELA and Fermi Anomalies,''
  arXiv:0911.5188 [hep-ph].
  %%CITATION = ARXIV:0911.5188;%%

\end{thebibliography}
\end{document}